# Polymorphism in Ruddlesden-Popper La$_3$Ni$_2$O$_7$: Discovery of a Hidden Phase with Distinctive Layer Stacking

Xinglong Chen,[1, *] Junjie Zhang,[1, #] A.S. Thind,[2] S. Sharma,[3] H. LaBollita,[3] G. Peterson,[1] H. Zheng,[1] D. Phelan,[1] A.S. Botana,[3] R.F. Klie,[2] and J.F. Mitchell[1, *]

[1]Materials Science Division, Argonne National Laboratory, Lemont, IL 60439, USA

[2]Department of Physics, University of Illinois Chicago, Chicago, IL 60607, USA

[3]Department of Physics and Astronomy, Arizona State University, Tempe, AZ 85218, USA

[*]Corresponding Authors: mitchell@anl.gov (J.F.M.), xinglong.chen@anl.gov (X.C.)

[#]Current Affiliation: Institute of Crystal Materials, Shandong University, Shandong, China

**Abstract:** We report the discovery of a novel form of Ruddlesden-Popper (RP) oxide, which stands as the first example of long-range, coherent polymorphism in this class of inorganic solids. Rather than the well-known, uniform stacking of perovskite blocks ubiquitously found in RP phases, this newly discovered polymorph of the bilayer RP phase La$_3$Ni$_2$O$_7$ adopts a novel stacking sequence in which single- and trilayer blocks of NiO$_6$ octahedra alternate in a "1313" sequence. Crystals of this new polymorph are described in space group *Cmmm*, although we note evidence for a competing *Imcm* variant. Transport measurements at ambient pressure reveal metallic character with evidence of a charge density wave transition with onset at $T \approx 134$ K, which lies intermediate between that of the standard "2222" polymorph of La$_3$Ni$_2$O$_7$ (space group *Amam*) and the trilayer RP phase, La$_4$Ni$_3$O$_{10}$. The discovery of such polymorphism could reverberate to the expansive range of science and applications that rely on RP materials, particularly the recently reported signatures of superconductivity with $T_c$ as high as 80 K above 14 GPa in bilayer La$_3$Ni$_2$O$_7$.

The Ruddlesden-Popper (RP) oxides, $A_{n+1}B_nO_{3n+1}$ (R=alkaline earth or rare earth, B=transition metal; $n$=1,2,3,…,∞) contain the structural motif of perovskite-like, $n$-layer slabs that are separated by double rocksalt layers. They are among the most well-studied inorganic solid state structure types.[1-9] Of particular importance is the notion that the dimensionality of their electronic structure evolves from quasi-two dimensional to three dimensional as $n$ increases. This dimensionality effect, along with distortion and tilting of oxygen octahedra within the perovskite slabs, strongly affects the behavior of a wide variety of materials, including solid oxide fuel cells, photocatalysts, metal-air battery cathodes, and a plethora of quantum materials that exhibit magnetism, metal-insulator transitions, superconductivity, etc. For example, the RP nickelates $La_{n+1}Ni_nO_{3n+1}$ evolve from insulating, charge- and spin-stripe ordered states for $n$=1,[10-15] to charge- and spin-density wave states for $n$=2 and $n$=3,[16-18] to a metallic, Pauli paramagnetic state for $n$=∞.[19-20] Indeed, high temperature superconductivity has recently been reported for $n$=2 and $n$=3 compounds subjected to pressure, with $La_3Ni_2O_7$ having a maximum reported $T_c \approx 80$ K (14 GPa)[21-25] and $La_4Ni_3O_{10} \approx 25$ K (38 GPa).[26-27] These discoveries are intriguing since the octahedral coordination and electronic configuration of Ni (formally, $d^{7.5}$, $Ni^{2.5+}$ for $La_3Ni_2O_7$ and $d^{7.33}$, $Ni^{2.67+}$ for $La_4Ni_3O_{10}$) are far-removed from the strongly Jahn-Teller distorted, $d^9$, $Cu^{2+}$ configuration of the superconducting cuprates.[28-29] These new results motivate the search for new RP compounds that could support superconductivity or competing electronic states.

In this Communication, we report the discovery of polymorphic RP phases in bulk crystals of $La_3Ni_2O_7$ established by combining single-crystal X-ray diffraction (SC-XRD) and transmission electron microscopy (TEM). The well-known $n$=2 RP $La_3Ni_2O_7$ polymorph (hereafter referred to as LNO-2222) adopts a uniform "2222" stacking of bilayers (Figure 1a). In contrast, the newly discovered polymorph (LNO-1313 hereafter) assumes a novel stacking sequence in which single-layer and trilayer blocks of NiO$_6$ octahedra alternate in a "1313" sequence (Figure 1b). Transport measurements on an LNO-1313 crystal reveal its metallic character and the signature of a charge density wave (CDW) at $T \approx 134$ K. First-principles calculations indicate that the electronic structure of LNO-1313 can be described as a superposition of the individual single- and trilayer subsystems.

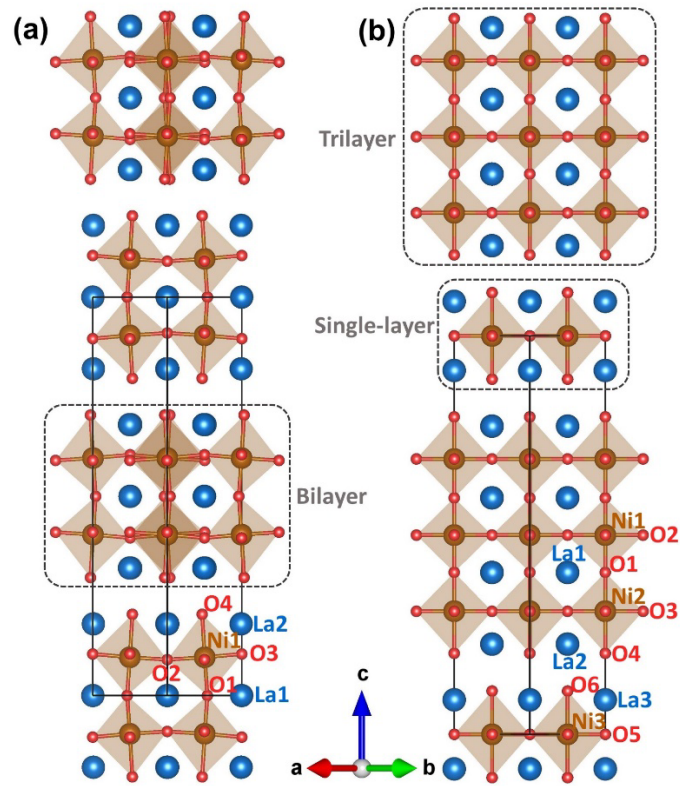

**Figure 1.** Polymorphic crystal structures of (a) LNO-2222 and (b) LNO-1313.

LNO-1313 single crystals were grown by a high oxygen pressure floating zone technique. During growth, competing phases include trilayer La$_4$Ni$_3$O$_{10}$, LNO-2222, and monolayer La$_2$NiO$_4$. Note that La$_4$Ni$_3$O$_{10}$ and La$_2$NiO$_4$ have been known as competing phases for LNO-2222 growth;[30] however, the 'hidden' LNO-1313 had not been identified until we discovered it serendipitously due to its distinctive single crystal X-ray diffraction (SC-XRD) pattern. We reproduced the growth of LNO-1313 crystals across several crystal boules, but at this stage the parameters that control the balance, selection, or distribution of phases during growth have not yet been fully identified. Since LNO-1313 has never been reported, despite the vast structural chemistry literature of polycrystalline bilayer RP phases, its appearance seems to be a unique consequence of growth from the melt.

Laboratory SC-XRD results reveal that LNO-1313 crystallizes in an orthorhombic space group, with reflection conditions consistent with *Cmmm* (No. 65). Remarkably, as shown in Figure 1b, rather than the expected uniform stacking of bilayers assembled from corner-shared $NiO_6$ octahedra, the LNO-1313 structure features an alternate stacking of single-layers and trilayers. In other words, the LNO-1313 polymorph can be viewed as a periodic, 1:1 ordered intergrowth of $La_2NiO_4$ and $La_4Ni_3O_{10}$. While incoherent stacking of layers has been reported in epitaxial films of RP phases,[31-32] to the best of our knowledge, no such bulk, long-range ordered polymorphic behavior has been reported in either oxide or non-oxide RP families.

The crystal data, structure refinement results, as well as bond lengths and bond angles of *Cmmm* LNO-1313 and *Amam* LNO-2222 are summarized in Table 1, with additional details in Tables S1-S9 in the supporting information (SI). While the unit cell volumes of these two phases are similar, the *a* and *b* axes of LNO-1313 are slightly longer than those of LNO-2222, while the *c* axis of LNO-1313 is slightly shorter. The Ni—Ni contact in LNO-1313 is 3.8566(3) Å in the *ab* plane and 3.8883(19) Å along *c*, both of which are shorter than those in LNO-2222 (3.8576(6) and 3.9421(12) Å, respectively). Rather than the buckled, ≈168° out-of-plane Ni1-O1-Ni1 angle found in LNO-2222, in *Cmmm* LNO-1313 the analogous angle (Ni1-O2-Ni2 in the trilayer block) is symmetry-constrained to 180°. Notably, this angle in trilayer RP $La_4Ni_3O_{10}$ *is* buckled.[30]

**Table 1 Crystal data, structure refinement, and selected bond lengths and angles for $La_3Ni_2O_7$ polymorphs.**

| Identification code | LNO-1313 | LNO-2222 |
| --- | --- | --- |
| Formula | $La_3Ni_2O_7$ | $La_3Ni_2O_7$ |
| Space group | *Cmmm* (No. 65) | *Amam* (No. 63) |
| $a$/Å | 5.4382(5) | 5.4053(6) |
| $b$/Å | 5.4700(5) | 5.4536(6) |
| $c$/Å | 20.3598(19) | 20.517(2) |
| Volume/Å$^3$ | 605.64(10) | 604.80(12) |
| Index ranges/h,k,l | [-6, 6], [-6, 6], [-25, 25] | [-7, 7], [-7, 7], [-29, 29] |
| Final $R$ indices [$I$>=2$\sigma$ ($I$)] | $R_1$ = 0.0238, $wR_2$ = 0.0458 | $R_1$ = 0.0202, $wR_2$ = 0.0527 |
| Largest diff. peak/hole /e Å$^{-3}$ | 0.99/-1.70 | 1.31/-1.14 |
| In-plane Ni-O/Å | Ni1-O2: 1.92832(12)<br>Ni2-O3: 1.92833(13)<br>Ni3-O5: 1.92832(13) | Ni1-O2: 1.9335(5)<br>Ni1-O3: 1.9183(6) |
| Out-of-plane Ni-O/Å | Ni1-O1: 1.902(13)<br>Ni2-O1: 1.986(13)<br>Ni2-O4: 2.150(13)<br>Ni3-O6: 2.211(12) | Ni1-O1: 1.9816(11)<br>Ni1-O4: 2.231(4) |
| In-plane Ni-O-Ni/° | Ni1-O2-Ni1: 180<br>Ni2-O3-Ni2: 179.7(5)<br>Ni3-O5-Ni3: 180 | Ni1-O2-Ni1: 172.0(2)<br>Ni1-O3-Ni1: 169.6(3) |
| Out-of-plane Ni-O-Ni/° | Ni1-O1-Ni2: 180 | Ni1-O1-Ni1: 168.1(3) |

As shown in Figure 2, the experimental diffraction patterns of LNO-1313 and LNO-2222 are quite distinct, especially along $c^*$. We describe both cells consistently (Table 1) with $a < b < c$. Consider first the ($hk$0) plane. Although LNO-1313 and LNO-2222 share the feature that prominent reflections are all observed at $h = 2n$, $k = 2n$, a set of weaker reflections (marked by open circles) can be used to distinguish their different centering modes, *C* and *A*, respectively. Additionally, in the (0$kl$) plane, the reflection density along $c^*$ in the of LNO-1313 is twice that of LNO-2222, clearly distinguishing the two polymorphs. For LNO-2222, the distinct reflections at $k = 2n +1$, $l = 2n + 1$ unambiguously demonstrate its *A* centering ($k + l = 2n$) rather than *F* centering ($k = 2n$, $l = 2n$ in (0$kl$)), which has been proposed for LNO-2222 by some authors.[33] A more detailed analysis is provided in the SI.

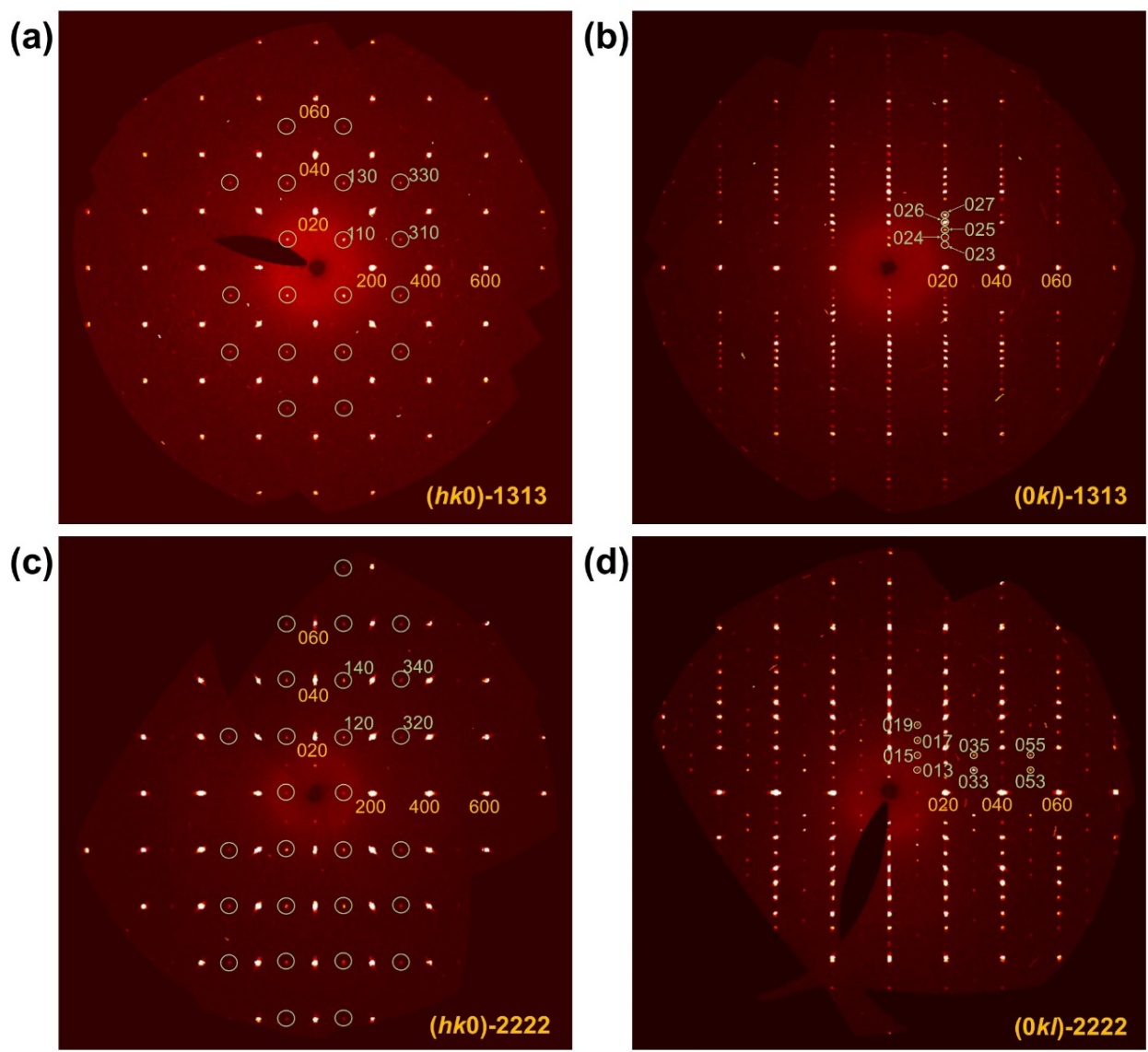

**Figure 2.** Precession images of LNO-1313 and LNO-2222 crystals. (a) (*hk*0) for LNO-1313; (b) (0*kl*) for LNO-1313; (c) (*hk*0) for LNO-2222; (d) (0*kl*) for LNO-2222.

We measured more than ten LNO-1313 specimens, all of which can be satisfactorily solved using the *Cmmm* model. However, in some specimens, weak half-integer reflections were observed in the (0*kl*) plane (see Figure S1) These reflections indicate the possibility of *c*-axis doubling and symmetry lowering to *Imcm* (alternate setting of *Imma*, No. 74), although this symmetry lowering preserves the "1313" stacking in the structure. Unfortunately, SC-XRD data alone preclude us from distinguishing whether these specimens are purely *Imcm* or a mixture of *Cmmm* and *Imcm* regions in the specimen. Further details are provided in the SI.

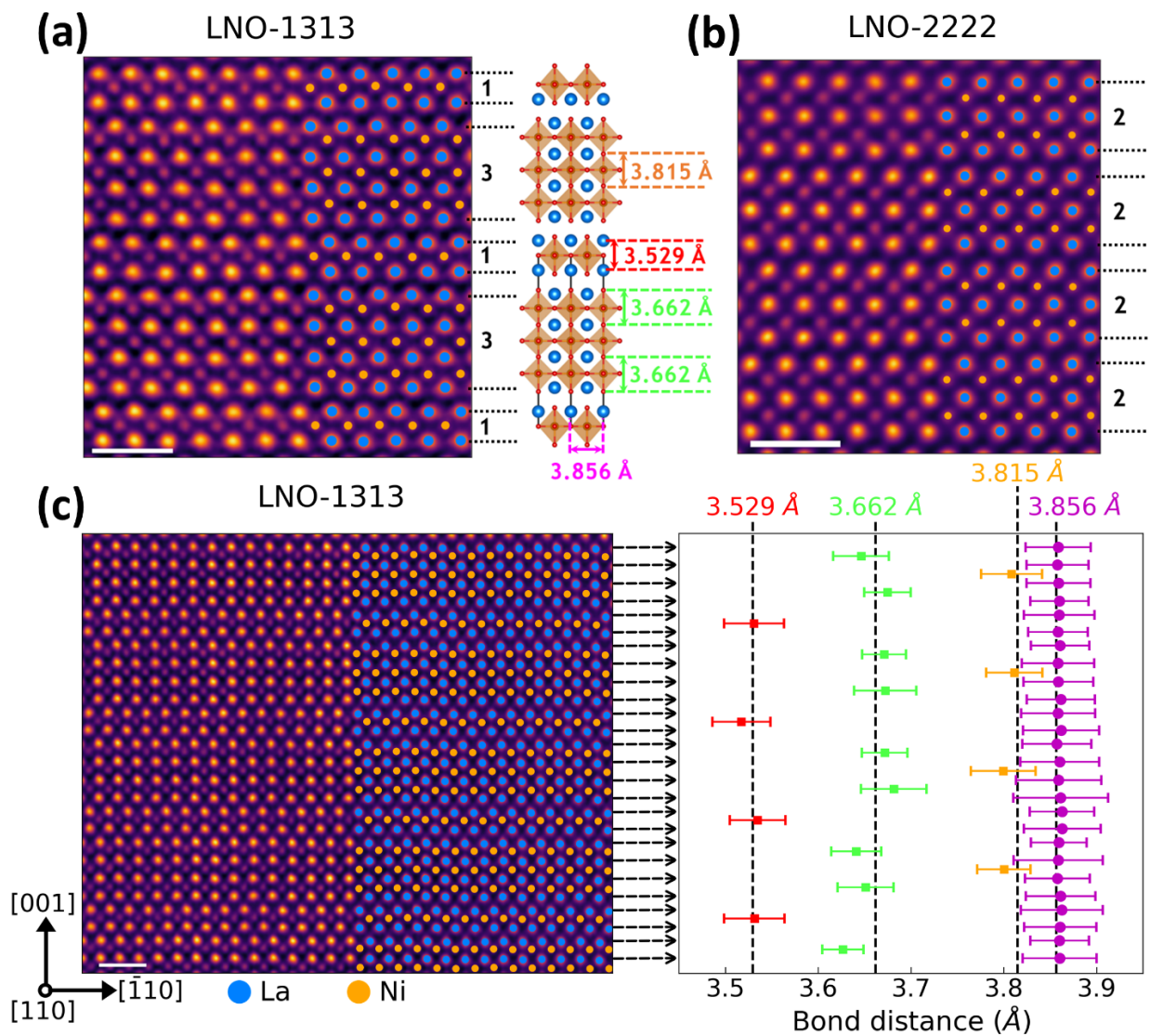

**Figure 3.** Atomic resolution STEM-HAADF images, with clearly distinguishable La and Ni atoms, showing the stacking sequence for (a) LNO-1313 and (b) LNO-2222 polymorphs along the [110] orientation. (c) Average layer-by-layer bond distance analysis of the La-La bonds for the STEM-HAADF image shown. The bond distances are defined in the atomic model from SC-XRD (same as Figure 1(b) as shown in (a)). The error bars correspond to the standard deviation. Scale bars correspond to 1 nm.

We used scanning transmission electron microscopy (STEM) to investigate the real space atomic structure and stacking sequence of LNO-1313 and LNO-2222 single crystals. Figures 3a and 3b show atomic-resolution high-angle annular dark-field (HAADF) images, with labeled La and Ni atomic columns, for LNO-1313 and LNO-2222 specimens, respectively, along the [110] crystallographic orientation. Each atomic column in a HAADF image can be distinguished based on the intensity profile because the HAADF intensity is approximately proportional to the squared atomic number ($\sim Z^2$) of the atomic column.[34] The La (Z = 57) atomic columns appear brightest followed by the Ni (Z = 28) atomic columns. Based on the cation sublattice, we can clearly visualize the "1313" and "2222" stacking sequences for the respective polymorphs, which are also labeled in Figure 3a and 3b.

Furthermore, we have investigated the layer-by-layer structure of the 1313 polymorph by analyzing the La-La bond distances across the monolayer and trilayer blocks. Figure 3c shows an atomic-resolution HAADF micrograph along with a comparison of the La-La bond distances, extracted from STEM-HAADF images and SC-XRD. Figure 3a shows the SC-XRD atomic model along with the bond distance definitions used for comparison with STEM image analysis. We find that the average layer-by-layer La-La bond distances calculated from our HAADF images agree well with the SC-XRD atomic model. As opposed to the uniform La-La bond lengths for the LNO-2222 polymorph, we clearly show that the La-La bond distances change across and within each stacking block for the LNO-1313 polymorph. For instance, the unique LNO-1313 layering sequence results in a shorter monolayer block, as well as shorter outer trilayer blocks in comparison to the LNO-2222 stacking. Whereas the middle trilayer block for the LNO-1313 polymorph is larger when compared with the uniform bilayer stacking of the LNO-2222 polymorph. Therefore, the STEM experiments corroborate the findings of our XRD results for this novel LNO-1313 polymorph. Further discussion on the morphology, composition, and defects for both LNO-1313 and LNO-2222 polymorphs can be found in the SI.

Figure 4 shows the normalized *ab*-plane resistivity of LNO-1313, LNO-2222 and La$_4$Ni$_3$O$_{10}$ single crystals between 1.8-200 K, all of which generally decrease with decreasing temperature, characteristic of a metal. For LNO-1313, a pronounced anomaly associated with a metal-to-metal transition is observed with onset $T_{MMT} \approx 134$ K. This electrical transport behavior resembles that of La$_4$Ni$_3$O$_{10}$ ($T_{MMT} \approx 139$ K) and the LNO-2222 polymorph of La$_3$Ni$_2$O$_7$ ($T_{MMT} \approx 122$ K). In these cases, the anomaly has been either shown to be a CDW by X-ray diffraction (La$_4$Ni$_3$O$_{10}$)[17] or speculated to be such (La$_3$Ni$_2$O$_7$)[16].

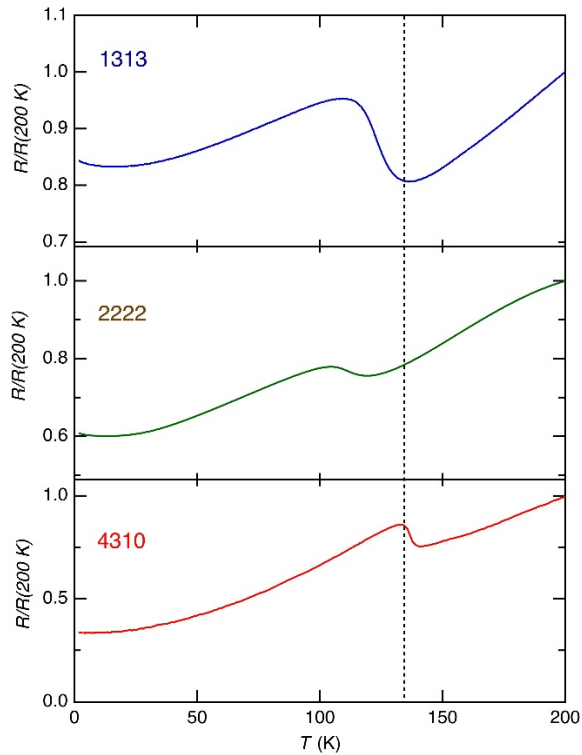

**Figure 4.** Electrical resistivity of LNO-1313, LNO-2222 and La$_4$Ni$_3$O$_{10}$ single crystals. The vertical dashed line marks the LNO-1313 transition onset determined by a tangent construction.

We studied the nonmagnetic electronic structure of the *Cmmm* structure using first-principles calculations. The band structure with band character plot of LNO-1313 (with $d^{7.5}$ filling) is shown in Figure 5. The most remarkable finding is that the band structure seems to be an electronic superposition of the single-layer ($La_2NiO_4$) and trilayer ($La_4Ni_3O_{10}$) subsystems at low-energy (although the superposition is not exact because of charge transfer from the formally $Ni^{8/3+}$ trilayer to the $Ni^{2+}$ single-layer to satisfy the $d^{7.5}$ electron count). This superposition is particularly noticeable when considering the dominant $e_g$ states around the Fermi level, where four Ni-$d_{x^2-y^2}$ and four Ni-$d_{z^2}$ bands can be seen (one from the single-layer Ni and three from the trilayer Ni atoms). For the $d_{z^2}$ states, the bonding, non-bonding, and anti-bonding bands from the Ni-$d_{z^2}$ triplet of the trilayer can be clearly appreciated.[34] This contrasts with the usual bonding-anti-bonding complex found in the LNO-2222 bilayer polymorph (Fermi surfaces and a comparison between the LNO-2222 and LNO-1313 electronic structures are shown in the SI). The fact that the electronic structure of $La_4Ni_3O_{10}$ underlies that of LNO-1313 is consistent with the similarities in the resistivity described above.

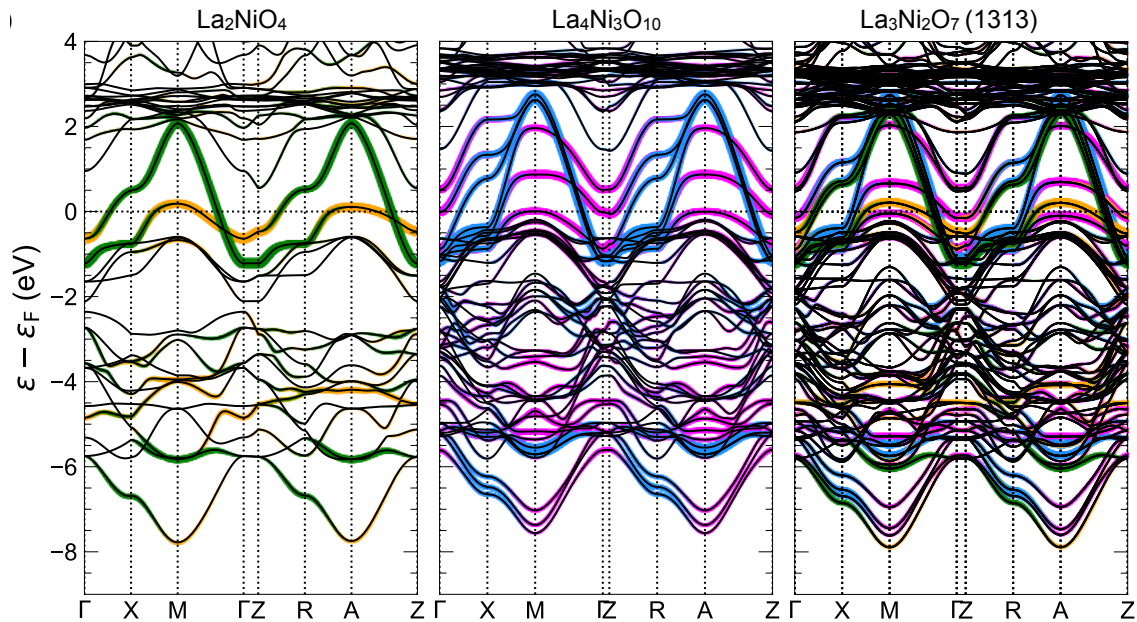

The report of superconductivity signatures in $La_3Ni_2O_7$ with $T_c$ up to 80 K under pressure represents a breakthrough in the quest for bulk nickelate superconductivity.[21] As mentioned earlier, both bilayer LNO-2222 and trilayer $La_4Ni_3O_{10}$ show CDW order in electrical transport measurements.[16-17] Under pressure, this CDW order is suppressed toward $T=0$, interrupted by the emergence of the superconducting transition. High pressure structural studies[21, 36] have found that the buckled Ni-O-Ni linkage joining octahedral layers in the bilayer nickelate straightens, coincident with the onset of superconductivity. Indeed, several authors argue that this structural response may be a prerequisite for superconductivity.[37-38] However, the proposed *Cmmm* structure of ambient-pressure, nonsuperconducting LNO-1313 contains 180º Ni-O-Ni linkages in the trilayer block, implying that this structural feature alone may be insufficient to stabilize the superconducting state. We would note that in the putative *Imcm* structure, which we have not been able to satisfactorily refine from SC-XRD data (see SI), buckling is possible and might respond to pressure.

Curiously, in bilayer nickelates a true zero-resistance state is not universally found, and, in certain samples, *no* superconducting transition is found.[39] Speculation that this inconsistency could signal filamentary superconductivity arising from a second unidentified phase in the bulk $La_3Ni_2O_7$ crystal has been strengthened by recent susceptibility measurements that reveal a rather low (approximately 1%) superconducting volume fraction in $La_3Ni_2O_7$ crystals under pressure.[40] While highly speculative, it is possible that one or more phases competing with LNO-2222 in the crystal growth, such as LNO-1313 (*Cmmm* and/or *Imcm*), could be a source of superconductivity. Measurements of LNO-1313 transport under pressure are essential to test this possibility and are underway.

In conclusion, we report a new form of alternating $n=1$ and $n=3$ perovskite slab layers in an ordered RP oxide phase, $La_3Ni_2O_7$. Two polymorphic stacking sequences—the known LNO-2222 and the new LNO-1313—are found during crystal growth at high $pO_2$. The discovery of these two polymorphs could fundamentally impact the rapidly evolving field of RP nickelate superconductivity. More broadly, the mere existence of such polymorphism tantalizes that similar polymorphism may be found in RP phases other than $La_3Ni_2O_7$ under suitable synthesis conditions. Such new compounds could harbor novel properties, with potential impact for the wide range of fundamental and applied science that rely on members of this expansive layered perovskite family.

**Acknowledgements:** Work at Argonne National Laboratory (crystal growth, structural determinations, transport measurements) was sponsored by the US Department of Energy, Office of Science, Basic Energy Sciences, Materials Science and Engineering Division. Work at UIC (microscopy) was supported by a grant from the National Science Foundation (NSF-DMR 22). Acquisition of the UIC JEOL ARM200CF was supported by an MRI-R2 grant from the National Science Foundation (DMR-0959470). The Gatan Continuum GIF acquisition at UIC was supported by an MRI grant from the National Science Foundation (DMR-1626065). This work made use of the ThermoFisher Helios 5CX (cryo) FIB-SEM instrument in the Electron Microscopy Core of UIC's Research Resources Center, which received support from UIC, Northwestern University and ARO (W911NF2110052). Work at ASU (theoretical calculations) was supported by NSF grant no. DMR-2045826 and the ASU Research Computing Center.

Supporting Information for

# Polymorphism in Ruddlesden-Popper La$_3$Ni$_2$O$_7$: Discovery of a Hidden Phase with Distinctive Layer Stacking

Xinglong Chen,[1,*] Junjie Zhang,[1,#] A.S. Thind,[2] S. Sharma,[3] H. LaBollita,[3] G. Peterson,[1] H. Zheng,[1] D. Phelan,[1] A.S. Botana,[3] R.F. Klie,[2] and J.F. Mitchell[1,*]

[1]Materials Science Division, Argonne National Laboratory, Lemont, IL 60439, USA

[2]Department of Physics, University of Illinois Chicago, Chicago, IL 60607, USA

[3]Department of Physics and Astronomy, Arizona State University, Tempe, AZ 85218, USA

*Corresponding Authors: mitchell@anl.gov (J.F.M.), xinglong.chen@anl.gov (X.C.)

#Current Affiliation: Institute of Crystal Materials, Shandong University, Shandong, China

# 1. Methods

## Crystal Growth

Crystals of LNO-1313 were grown by a high-pressure floating zone method in a vertical optical-image furnace (150-bar Model HKZ, SciDre GmbH, Dresden). Stoichiometric mixtures of $La_2O_3$ (pretreated at 1000 °C) and NiO were ground and fired at 1050 °C for 3 days with several intermediate grindings. These precursor powders were hydrostatically pressed into a rod and sintered at 1400 °C for 12-24 h. Crystals of LNO-1313 were grown directly from the sintered rods at in 100% $O_2$ at a pressure of 13-16 bars. During the crystal growth, the traveling rate was 3-4mm/h, and the feed rod and seed were counter-rotated at a rate in the range of 15-27 rpm. LNO-1313 crystals were obtained from some sections of the grown boules. Crystals of LNO-2222 were also grown under similar conditions. The two kinds of crystals were identified by X-ray diffraction.

## Single-Crystal X-ray Structure Determination

Single-crystal X-ray diffraction (SC-XRD) data were collected at room temperature using a Bruker D8 diffractometer equipped with APEX2 area detector and Mo K$\alpha$ radiation ($\lambda$ = 0.71073 Å). Data integration, cell refinement and numerical absorption corrections were performed by the SAINT program and SADABS program in APEX3 software.[1,2] The precession images were also synthesized in APEX3. The structure was solved by Olex2 using direct methods with the XS structure solution program and refined with full-matrix least-squares methods on F$^2$ by the XL refinement package.[3,4] PLATON was applied to check for possible missing symmetry.[5]

**Electron Microscopy**

The samples for scanning transmission electron microscopy (STEM) experiments were prepared, from bulk LNO-1313 and LNO-2222 single crystals, using a Thermo Fischer Scientific Helios 5 CX focused-ion beam (FIB)/scanning electron microscope (SEM) DualBeam system at the University of Illinois Chicago. To avoid ion-beam damage during sample preparation, a protective coating of W was deposited on the surface of the crystals. The electron transparent samples for STEM experiments were obtained by carrying out the final thinning with an ion beam energy of 1 kV. The samples were mounted on a Mo FIB grid. SEM experiments on the bulk LNO-1313 and LNO-2222 single crystals were performed simultaneously during the sample preparation. SEM imaging and energy-dispersive X-ray spectroscopy (EDS) were performed with an accelerating voltage of 30 kV and a beam current of 1.4 nA.

STEM imaging and spectroscopy experiments were performed at the University of Illinois Chicago using an aberration-corrected JEOL JEM-ARM200CF microscope, operated at 200 kV. The microscope is equipped with a cold-field emission electron gun and a CEOS aberration corrector. A probe convergence angle of 30 mrad was used for high-angle annular dark-field (HAADF) imaging. The inner and outer acceptance angles for the annular detector were set to 68 mrad and at least 174.5 mrad respectively. The atomic-resolution HAADF images were acquired sequentially, where multiple frames (10-30) were aligned and integrated upon the removal of scan distortions. The atomic positions in the HAADF images were initially estimated using relative intensity and were then further refined upon using 2D Gaussian fitting. The refined atomic positions were used for bond distance analysis shown in Figure 3.

### Computational Studies

The electronic structure of LNO-1313 was investigated via density-functional theory (DFT)-based calculations using the all-electron, full potential code WIEN2k within the local-density approximation (LDA).[6] A dense 17x17x3 $k$-mesh was used for sampling the Brillouin zone. The basis size is set by $RK_{max}$ = 7 and muffin-tin radii (in a.u.): 2.35, 1.90, 1.69 for La, Ni, O, respectively.

### Electrical Resistivity

Electrical resistivity of LNO-1313, LNO-2222, and $La_4Ni_3O_{10}$ single crystals was measured using a four-terminal method in a Quantum Design PPMS in the temperature range of 1.8-300 K.

## 2. Results and Discussion

### Crystal Structure of LNO-1313

The structure difference between LNO-1313 (*Cmmm*) and LNO-2222 (*Amam*) is clearly reflected in precession images of the experimental diffraction pattern. To simplify the comparison, the *Amam* setting is chosen for LNO-2222, thus the two structures have the same order of lattice constants, $a < b < c$. Fig. 2 shows the synthesized precession images of the ($hk0$) and ($0kl$) planes in reciprocal space for LNO-1313 and LNO-2222 crystals. As seen in Fig. 2a and 2c, the common feature in ($hk0$) for LNO-1313 and LNO-2222 is that the prominent reflections are all observed at $h = 2n$, $k = 2n$. However, the weak reflections (marked by open circles) can be used to distinguish the two phases owing to their different symmetry. Specifically, for LNO-1313, the weak reflections only appear at $h = 2n + 1$, $k = 2n + 1$, together with the subset of strong reflections indicative of the *C*-centering ($h + k = 2n$). In contrast, for LNO-2222 the weak reflections show up at $h = 2n + 1$ and $k = 2n$ (required by the *A*-centering ($k + l = 2n$)).

The (0*kl*) plane is likewise diagnostic. By comparing Figure 2b and 2d, one finds that the reflection density along $c^*$ in the (0*kl*) plane of 1313 is twice as that of LNO-2222. Further analysis reveals that in the (0*kl*) plane of LNO-1313, all reflections are observed at $k = 2n$ arising from the *C* centering condition that $h + k = 2n$, while $l$ can be either even or odd. In contrast, prominent reflections in (0*kl*) of LNO-2222 are seen when $k = 2n$, $l = 2n$, however, distinct reflections are observed at $k = 2n +1$, $l = 2n + 1$, which together imply *A* centering ($k + l = 2n$) rather than *F* centering ($k = 2n$, $l = 2n$ in (0*kl*)).

We note that in some specimens, additional weak, half-integer reflections are observed along $c^*$ in the (0*kl*) plane (see Fig. S1), even though the structure of these specimens can be well determined and refined using the *Cmmm* model. These additional weak features indicate the possibility of *c*-axis doubling and possible symmetry lowering in these samples. The symmetry lowering will not change the "1313" stacking sequence; however, the lower symmetry permits tilting of the NiO$_6$ octahedra in the trilayer block, thus the out of plane Ni-O-Ni angle need not be 180° in the corresponding structure model.

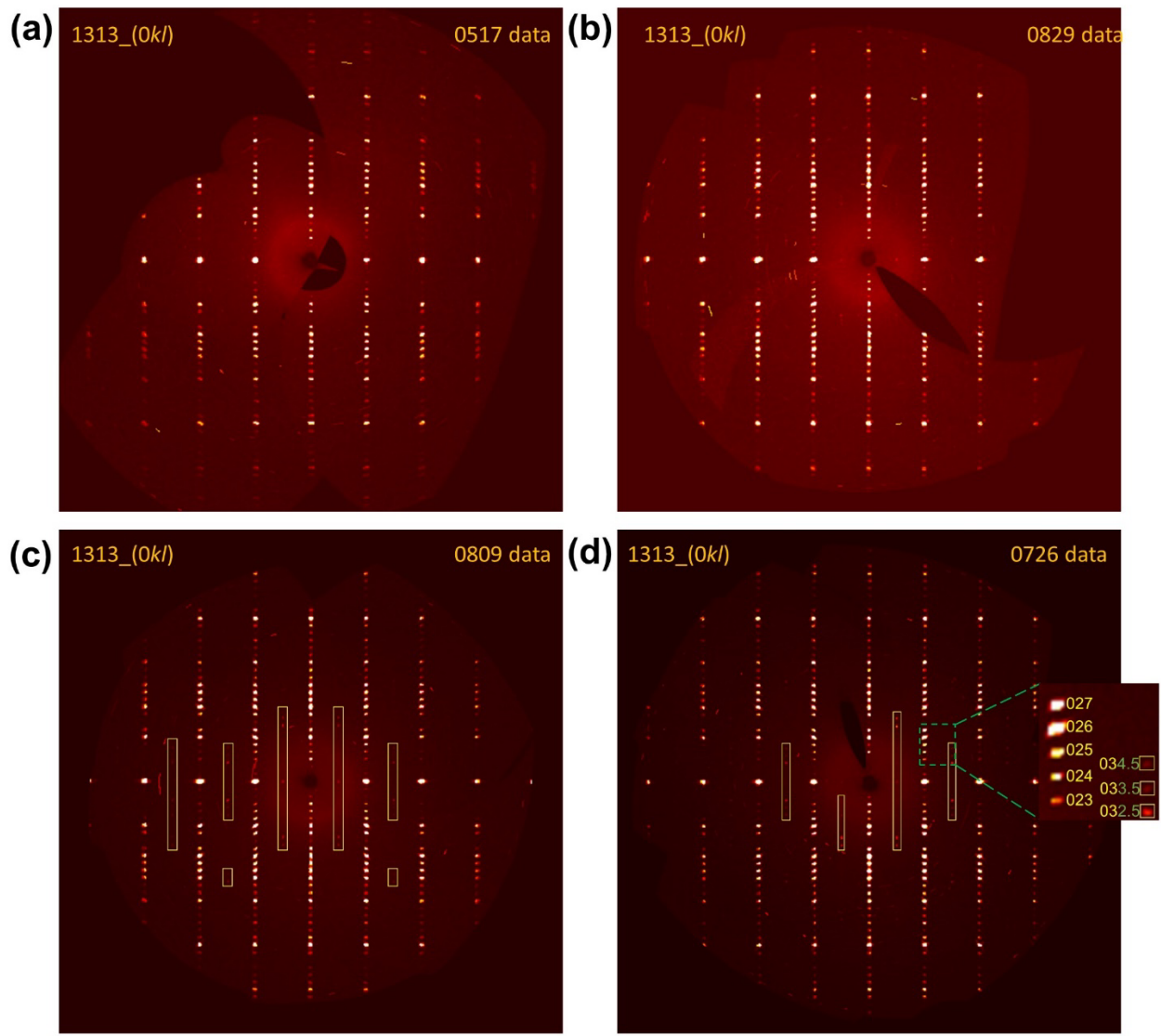

**Fig. S1** Synthesized precession images of (*0kl*) for four LNO-1313 samples. (a) and (b) do not have half-integer reflections in (*0kl*), (c) and (d) contain weak half-integer reflections (some of them are marked by open rectangles).

We analyzed the XRD data for these samples taking the weak reflections (marked by open rectangles in Fig. S1(c) and (d)) into consideration and found that *Imcm* with lattice constants ordered as $a < b < c$ (alternate setting of *Imma*, No. 74) best fit the reflection conditions. The crystal data, structure refinement results, and atomic coordinates are listed in Tables S10 and S11. Notably, the final refinement results based on *Imcm* are unsatisfactory, with several Ni and O atoms having negative anisotropic displacement parameters (ADP). The possible reasons for the poor refinement result based on the *Imcm* model in these specimens could be: (1) the limitation of laboratory X-ray for detecting accurate O positions in 1313. Inspection indicates that the intensity of most of these additional reflections (i.e., those not explained by *Cmmm*) in 1313 samples are much weaker than that of the nearby prominent reflections attributed to *Cmmm*. (2) The *Cmmm* and *Imcm* phases coexist in certain LNO-1313 crystals, and the *Imcm* phase has a relatively small volume fraction in the specimens that we measured. Since (1) the two phases share many common reflections, (2) structural models of the two phases do not exist, and (3) the volume fractions are also unknown, structural refinement presents an underdetermined problem given only a single X-ray dataset. Further investigations are needed to address this issue.

**Electron Microscopy**

We have performed SEM imaging and SEM energy-dispersive X-ray spectroscopy (EDS) experiments to investigate the morphology and composition of the bulk LNO-1313 and LNO-2222 single crystals. The electron images in Fig. S2(a) and S2(b) show the bulk single crystals for the LNO-1313 and LNO-2222 polymorphs, respectively, attached to carbon tape. The crystal size for the LNO-1313 polymorph varies from 200 μm to 400 μm, whereas the crystal size for the LNO-2222 polymorph was observed to be about 500 μm. The crystals for both polymorphs have a high surface roughness without preferred high symmetry surface terminations. The relative chemical composition of La and Ni for both polymorphs (59.4% La and 40.6% Ni for the LNO-1313 polymorph and 57.2 % La and 42.8% Ni for the LNO-2222 polymorph) is consistent with the expected relative composition of 60% La and 40% Ni.

Apart from the stacking sequence viewed along [110] (as shown in Figure 3), we also show the "1313" stacking sequence along [100] in Fig. S3(a). Overall, we observed the specimen containing the LNO-1313 polymorph to be high quality with a very low density of stacking faults. The typical stacking fault for the LNO-1313 specimens that we observed is an occasional intergrowth of a trilayer instead of a monolayer as shown in Fig. S3(b). For the specimens containing the LNO-2222 polymorph, we observed a higher density of stacking faults. The most prominent stacking fault for the LNO-2222 specimens is observed to be the growth of even numbered trilayer stacks instead of bilayer stacks.

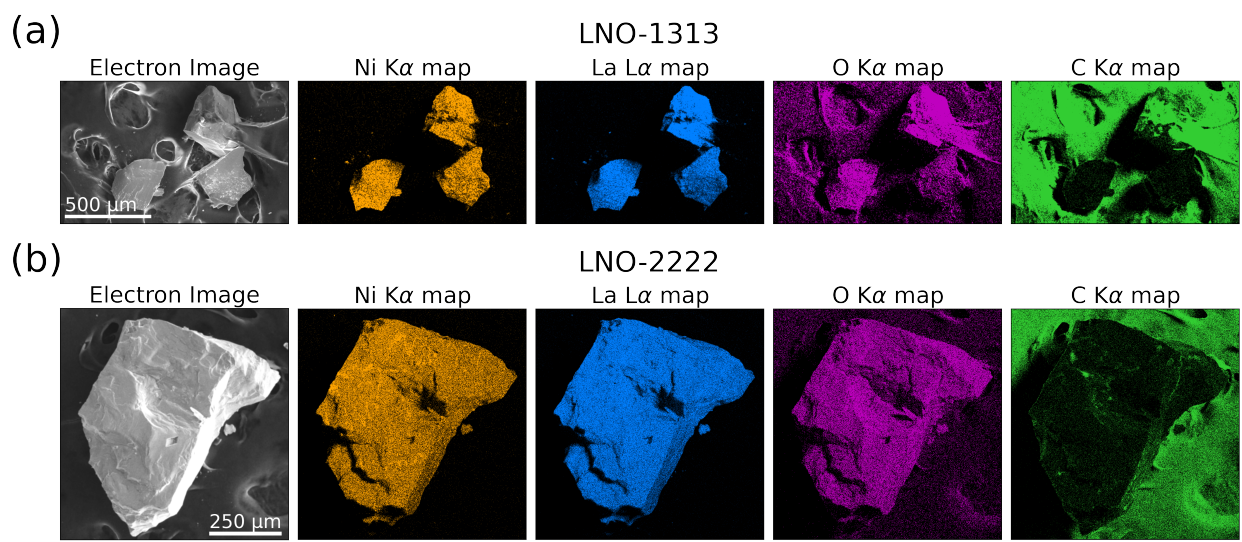

**Fig. S2.** SEM images and SEM-EDS chemical maps of Ni, La, O, and C for the (a) LNO-1313 and (b) LNO-2222 crystals.

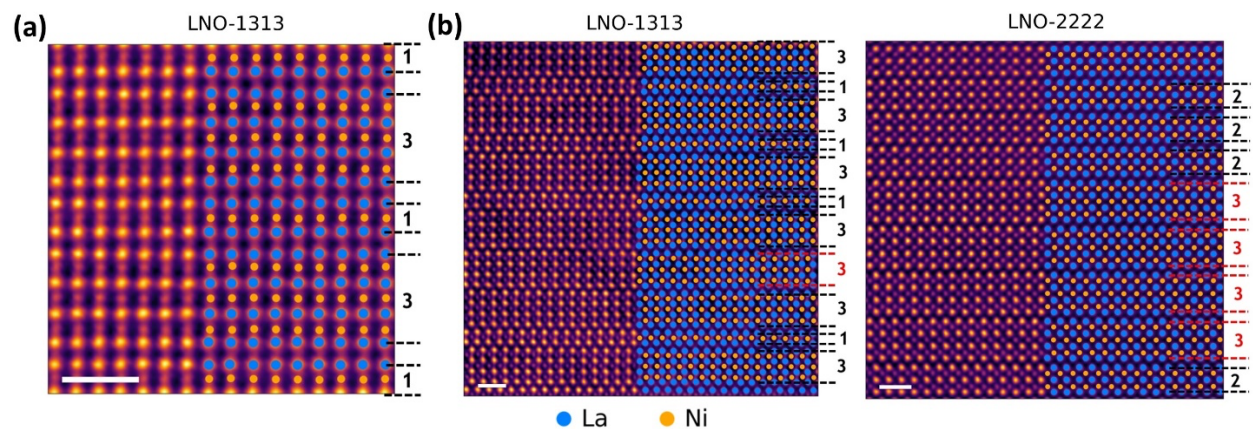

**Fig. S3.** (a) Atomic resolution HAADF image showing the "1313" stacking along [100]. (b) Atomic resolution HAADF images showing stacking faults (marked in red) for the LNO-1313 and LNO-2222 polymorphs. Scale bars correspond to 1 nm.

**Theoretical Calculations**

Additional details of the calculated electronic band structure of *Cmmm* LNO-1313 are provided in Fig. S4. Fig. S4(a) shows the structure of LNO-1313 with labels for each Ni in its alternating single-layer and trilayer along the *c*-axis. Fig. S4(b) shows the LDA band structure with Ni-$e_g$ ($d_{x^2-y^2}$, $d_{z^2}$) orbital character highlighted. Four Ni-$d_{x^2-y^2}$ and four Ni-$d_{z^2}$ bands can be observed near the Fermi level (one from the single layer and three from the trilayer). As such, the band structure reveals an electronic superposition of the monolayer and trilayer subsystems at low energy, as explained in the main text. The corresponding Fermi surface is analogous to that of other RP nickelates: large hole pockets of $d_{x^2-y^2}$ character (resembling cuprate-like arcs), that are centered at the corner of the Brillouin zone (the number of these pockets is equal to the number of inequivalent Ni atoms). In addition to the $d_{x^2-y^2}$ hole-like sheets, extra electronlike pockets appear at the zone center of mixed $d_{x^2-y^2}+d_{z^2}$ character or pure $d_{z^2}$ character (see Fig. S4(c)). The atom-resolved density of states (DOS) (Fig. 4(d)) reveals the low-energy states are an admixture of Ni(*3d*) and O(*2p*) states. The La(*5d*) states are not present at the Fermi level, which can be understood by the much larger Ni valence (Ni$^{2.5+}$) compared to the reduced RP nickelates.

An important quantity in transition-metal compounds, especially in the context of high-$T_c$ superconductivity, is the degree of hybridization between the transition-metal (Ni) and ligands (O). Here, we calculate the charge-transfer energy, $\Delta = \varepsilon_d - \varepsilon_p$ as the center-of-mass difference between the Ni(*3d*) band and O(*2p*) bands to be $\Delta = 3$ eV, which is larger than typical cuprate values (1-2 eV),[7] but significantly reduced from the reduced RP (e.g., LaNiO$_2$) nickelates (~4 eV).[8]

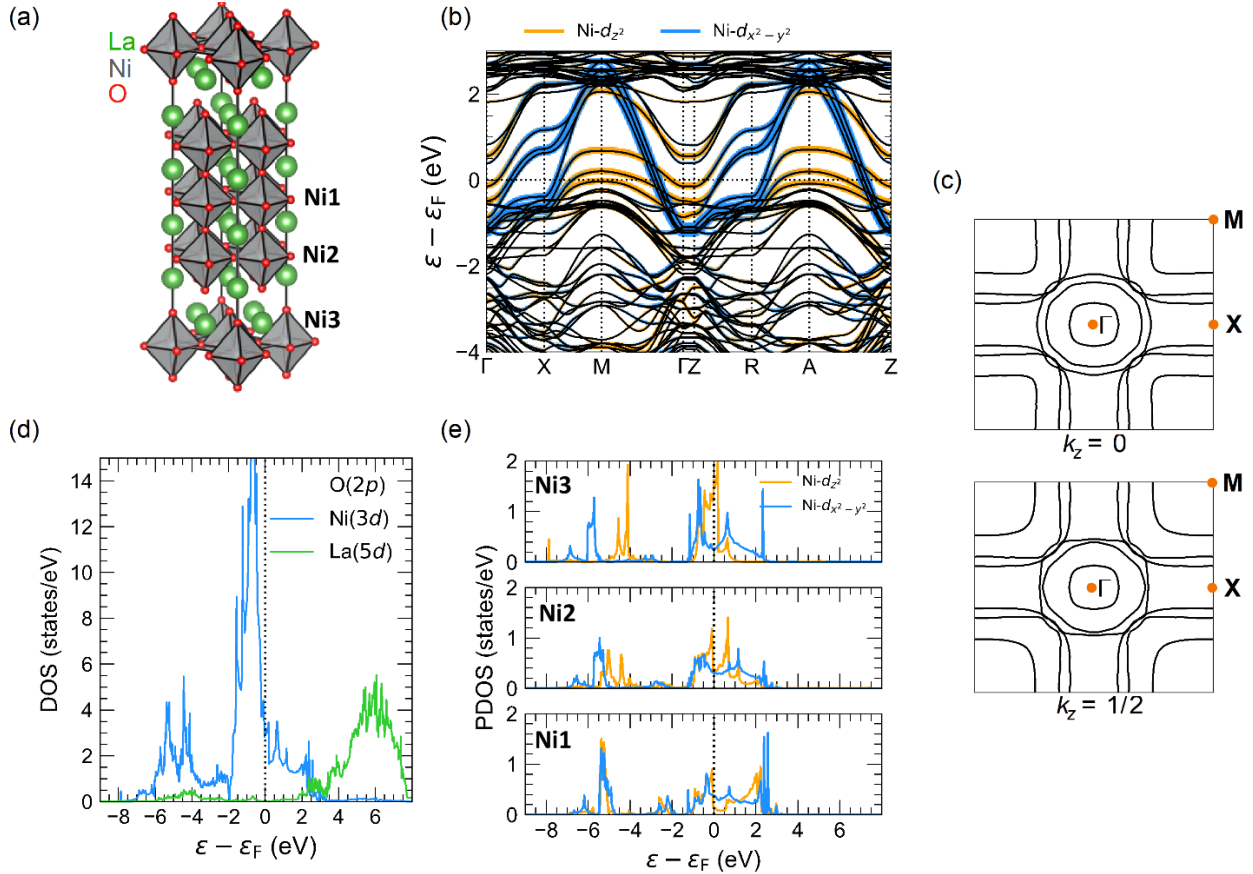

**Fig. S4.** LDA electronic structure of LNO-1313 (*Cmmm*). (a) Crystal structure highlighting the three inequivalent Ni atoms: Ni3 corresponding to the single layer, Ni2 and Ni1 corresponding to the outer and inner inequivalent Ni sites within the trilayer block. La, Ni, O atoms are denoted by green, grey, and red spheres, respectively. (b) Orbital-resolved band structure along high-symmetry lines highlighting the Ni-$d_{z^2}$ (orange) and Ni-$d_{x^2-y^2}$ (blue) orbital weights. (c) Fermi surfaces at two different $k_z$ momentum cuts: $k_z = 0$ (top) and $k_z = ½$ (bottom). (d) Atom-resolved density of states (DOS) in a wide energy window. (e) Site-resolved Ni-$e_g$ partial density of states (PDOS).

The site- and orbital-resolved density of states are shown in Fig. S4(e). Within the single-layer (Ni3), the low-energy states are dominated by hybridized Ni-$e_g$ orbitals with a large-peak around the Fermi level due to the flat Ni-$d_{z^2}$ band at the M point. Within the trilayer, there are two inequivalent Ni sites, Ni2 and Ni1, that correspond to the outer and inner layers, respectively (see Fig. S4(a)). The PDOS reveals slightly different low-energy physics between the two layers; this can be understood from a molecular-orbital picture, where the trilayer block forms a trimer along the *c*-axis with bonding (even symmetry), non-bonding (odd symmetry), and anti-bonding (even symmetry) states.

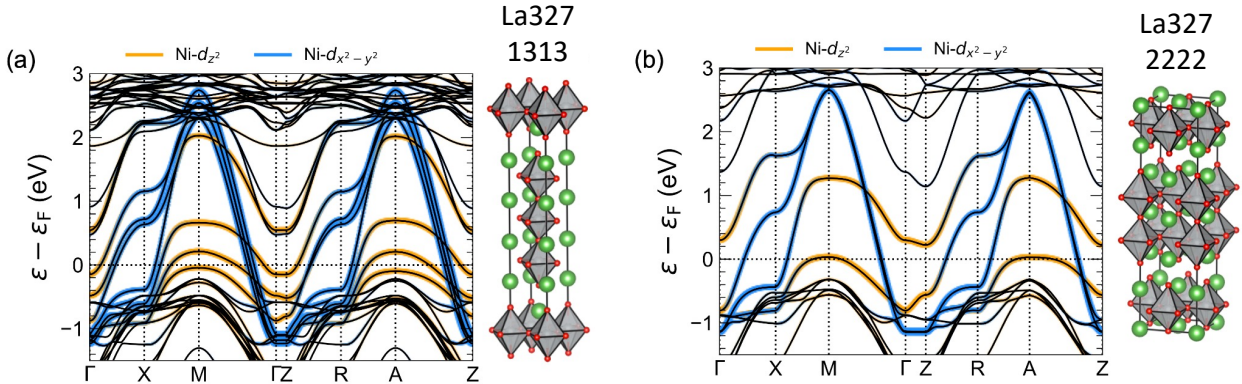

**Fig. S5.** Orbital-resolved bands structure along high-symmetry lines for (a) LNO-1313(*Cmmm*) and (b) LNO-2222 (*Fmmm*) (we plot the band structure of the LNO-2222 polymorph in *Fmmm* symmetry to be able to draw direct comparisons to the LNO-1313 counterpart).

Interestingly, when comparing the electronic structure of LNO-1313 to LNO-2222 (see Fig. S5), while many of the features are the same—displaying strongly hybridized Ni-$e_g$ states at low-energy with $d_{x^2-y^2}$ bands of approximately the same bandwidth—there are subtle differences arising from the underlying structural makeups. One of the more noticeable differences is in the different splittings of the $d_{z^2}$ states in terms of molecular orbitals (that have been deemed to be crucial for superconductivity in the 2222 counterpart under pressure[9]): a bonding-antibonding orbital complex forms in LNO-222 while in the LNO-1313 polymorph a bonding-nonbonding-antibonding complex is inherited from the trilayer with an extra $d_{z^2}$ band from the single-layer block. Further, the LNO-1313 polymorph displays a more marked 2D character for the electronic structure.

# 3. Tables of Crystallographic Data

**Table S1 Crystal data and structure refinement results for LNO-1313 and LNO-2222.**

| Identification code | LNO-1313 | LNO-2222 |
|---|---|---|
| Empirical formula | $La_3Ni_2O_7$ | $La_3Ni_2O_7$ |
| Formula weight | 646.15 | 646.15 |
| Temperature/K | 295 | 295 |
| Crystal system | orthorhombic | orthorhombic |
| Space group | *Cmmm* | *Amam* |
| $a$/Å | 5.4382(5) | 5.4053(6) |
| $b$/Å | 5.4700(5) | 5.4536(6) |
| $c$/Å | 20.3598(19) | 20.517(2) |
| $\alpha$/° | 90 | 90 |
| $\beta$/° | 90 | 90 |
| $\gamma$/° | 90 | 90 |
| Volume/Å$^3$ | 605.64(10) | 604.80(12) |
| Z | 4 | 4 |
| $\rho_{calc}$g/cm$^3$ | 7.086 | 7.096 |
| $\mu$/mm$^{-1}$ | 26.793 | 26.830 |
| F(000) | 1132.0 | 1132.0 |
| Crystal size/mm$^3$ | 0.03 × 0.028 × 0.02 | 0.085 × 0.076 × 0.03 |
| Radiation | Mo K$\alpha$ ($\lambda$ = 0.71073) | Mo K$\alpha$ ($\lambda$ = 0.71073) |
| 2$\Theta$ range for data collection/° | 2 to 52.784 | 7.732 to 60.934 |
| Index ranges | -6 ≤ h ≤ 6, -6 ≤ k ≤ 6, -25 ≤ l ≤ 25 | -7 ≤ h ≤ 7, -7 ≤ k ≤ 7, -29 ≤ l ≤ 29 |
| Reflections collected | 4903 | 5833 |
| Independent reflections | 395 [$R_{int}$ = 0.0462, $R_{sigma}$ = 0.0217] | 514 [$R_{int}$ = 0.0283, $R_{sigma}$ = 0.0144] |
| Completeness | 100% | 100% |
| Refinement method | Full-matrix least-squares on $F^2$ | Full-matrix least-squares on $F^2$ |
| Absorption correction | Numerical | Numerical |
| Goodness-of-fit on $F^2$ | 1.283 | 1.250 |
| Final $R$ indexes [$I$>=2σ ($I$)] | $R_1$ = 0.0238, $wR_2$ = 0.0458 | $R_1$ = 0.0202, $wR_2$ = 0.0527 |
| Final $R$ indexes [all data] | $R_1$ = 0.0313, $wR_2$ = 0.0486 | $R_1$ = 0.0219, $wR_2$ = 0.0535 |
| Largest diff. peak/hole / e Å$^{-3}$ | 0.99/-1.70 | 1.31/-1.14 |

**Table S2** Fractional atomic coordinates (×10⁴) and equivalent isotropic displacement parameters (Å²×10³) for LNO-1313 in *Cmmm*. $U_{eq}$ is defined as 1/3 of the trace of the orthogonalized $\underline{U}_{ij}$ tensor.

| Atom | Wyck. | x | y | z | $U_{eq}$ |
|---|---|---|---|---|---|
| La1 | 4l | 5000 | 0 | 5936.8(5) | 9.4(3) |
| La2 | 4l | 5000 | 0 | 7735.5(4) | 7.5(3) |
| La3 | 4k | 5000 | 5000 | 9133.2(5) | 7.9(3) |
| Ni1 | 2b | 5000 | 5000 | 5000 | 3.6(6) |
| Ni2 | 4k | 5000 | 5000 | 6909.9(9) | 5.7(5) |
| Ni3 | 2b | 0 | 5000 | 10000 | 7.7(7) |
| O1 | 4k | 5000 | 5000 | 4066(6) | 37(4) |
| O2 | 4f | 7500 | 2500 | 5000 | 40(4) |
| O3 | 8m | 7500 | 2500 | 6912(4) | 15(2) |
| O4 | 4k | 5000 | 5000 | 7966(6) | 27(4) |
| O5 | 4e | 2500 | 2500 | 10000 | 11(3) |
| O6 | 4l | 0 | 5000 | 8914(6) | 27(4) |

**Table S3** Anisotropic Displacement Parameters (Å²×10³) for LNO-1313 (*Cmmm*). The Anisotropic displacement factor exponent takes the form: $-2\pi^2[h^2a^{*2}U_{11}+2hka^*b^*U_{12}+\ldots]$.

| Atom | $U_{11}$ | $U_{22}$ | $U_{33}$ | $U_{23}$ | $U_{13}$ | $U_{12}$ |
|---|---|---|---|---|---|---|
| La1 | 13.6(5) | 8.1(5) | 6.4(5) | 0 | 0 | 0 |
| La2 | 8.0(5) | 8.5(5) | 6.1(5) | 0 | 0 | 0 |
| La3 | 8.9(5) | 8.6(5) | 6.1(5) | 0 | 0 | 0 |
| Ni1 | 5.3(16) | 3.6(14) | 1.8(12) | 0 | 0 | 0 |
| Ni2 | 6.9(12) | 5.2(11) | 4.9(9) | 0 | 0 | 0 |
| Ni3 | 7.5(17) | 4.2(15) | 11.5(14) | 0 | 0 | 0 |
| O1 | 38(9) | 70(10) | 5(6) | 0 | 0 | 0 |
| O2 | 41(11) | 33(10) | 46(9) | 0 | 0 | 33(8) |
| O3 | 3(5) | 5(5) | 37(5) | 0 | 0 | 2(3) |
| O4 | 16(8) | 53(10) | 13(6) | 0 | 0 | 0 |
| O5 | 4(7) | 7(7) | 22(6) | 0 | 0 | 1(5) |
| O6 | 27(8) | 45(9) | 10(6) | 0 | 0 | 0 |

**Table S4 Selected bond lengths for LNO-1313.**

| Atom | Atom | Length/Å | Atom | Atom | Length/Å |
|---|---|---|---|---|---|
| La1 | O1[1] | 2.7191(3) | La3 | O5 | 2.6139(7) |
| La1 | O1[2] | 2.7350(3) | La3 | O5[15] | 2.6139(7) |
| La1 | O1[3] | 2.7350(3) | La3 | O6 | 2.7555(19) |
| La1 | O1[4] | 2.7191(3) | La3 | O6[10] | 2.7712(19) |
| La1 | O2[5] | 2.7123(7) | La3 | O6[13] | 2.7712(19) |
| La1 | O2[6] | 2.7123(7) | La3 | O6[12] | 2.7555(19) |
| La1 | O2[7] | 2.7123(7) | Ni1 | O1 | 1.902(13) |
| La1 | O2 | 2.7123(7) | Ni1 | O1[2] | 1.902(13) |
| La1 | O3[8] | 2.768(6) | Ni1 | O2 | 1.92832(12) |
| La1 | O3[5] | 2.768(6) | Ni1 | O2[17] | 1.92832(13) |
| La1 | O3 | 2.768(6) | Ni1 | O2[18] | 1.92832(13) |
| La1 | O3[9] | 2.768(6) | Ni1 | O2[7] | 1.92832(13) |
| La2 | O3[5] | 2.555(5) | Ni2 | O1[2] | 1.986(13) |
| La2 | O3[9] | 2.555(5) | Ni2 | O3 | 1.92833(13) |
| La2 | O3[8] | 2.555(5) | Ni2 | O3[17] | 1.92833(13) |
| La2 | O3 | 2.555(5) | Ni2 | O3[9] | 1.92833(13) |
| La2 | O4[5] | 2.759(2) | Ni2 | O3[19] | 1.92833(14) |
| La2 | O4[10] | 2.759(2) | Ni2 | O4 | 2.150(13) |
| La2 | O4 | 2.775(2) | Ni3 | O5[21] | 1.92832(13) |
| La2 | O4[11] | 2.775(2) | Ni3 | O5[17] | 1.92832(13) |
| La2 | O6[10] | 2.400(12) | Ni3 | O5 | 1.92832(13) |
| La3 | O4 | 2.377(12) | Ni3 | O5[14] | 1.92832(13) |
| La3 | O5[14] | 2.6139(7) | Ni3 | O6[20] | 2.211(12) |
| La3 | O5[13] | 2.6139(7) | Ni3 | O6 | 2.211(12) |

[1] 3/2-X,1/2-Y,1-Z; [2] 1-X,1-Y,1-Z; [3] 1-X,-Y,1-Z; [4] 1/2-X,1/2-Y,1-Z; [5] -1/2+X,-1/2+Y,+Z; [6] +X,-Y,1-Z; [7] -1/2+X,1/2-Y,1-Z; [8] 3/2-X,-1/2+Y,+Z; [9] 1-X,+Y,+Z; [10] 1/2-X,-1/2+Y,+Z; [11] +X,-1+Y,+Z; [12] 1+X,+Y,+Z; [13] 1/2+X,1/2+Y,+Z; [14] +X,1-Y,2-Z; [15] 1/2+X,1/2-Y,2-Z; [16] +X,1+Y,+Z; [17] -1/2+X,1/2+Y,+Z; [18] +X,1-Y,1-Z; [19] 3/2-X,1/2+Y,+Z; [20] -X,1-Y,2-Z; [21] -1/2+X,1/2-Y,2-Z;

**Table S5 Selected bond angles for LNO-1313.**

| Atom | Atom | Atom | Angle/° | Atom | Atom | Atom | Angle/° |
|---|---|---|---|---|---|---|---|
| O1[1] | La1 | O1[2] | 90.000(3) | O6 | La3 | O6[12] | 88.50(8) |
| O1[1] | La1 | O1[3] | 90.000(2) | O1 | Ni1 | O1[2] | 180 |
| O1[4] | La1 | O1[2] | 90.000(2) | O1[2] | Ni1 | O2 | 90 |
| O1[2] | La1 | O1[3] | 179.8(6) | O1[2] | Ni1 | O2[17] | 90 |
| O1[4] | La1 | O1[3] | 90.000(3) | O1 | Ni1 | O2[8] | 90.000(1) |
| O1[1] | La1 | O1[4] | 179.8(6) | O1 | Ni1 | O2[18] | 90.000(1) |
| O1[1] | La1 | O3[5] | 119.5(2) | O1 | Ni1 | O2[17] | 90 |
| O1[2] | La1 | O3[5] | 119.7(2) | O1[2] | Ni1 | O2[18] | 90.000(1) |
| O1[1] | La1 | O3[6] | 60.7(2) | O1[2] | Ni1 | O2[8] | 90.000(1) |
| O1[3] | La1 | O3[6] | 60.5(2) | O1 | Ni1 | O2 | 90 |
| O1[3] | La1 | O3[5] | 60.5(2) | O2[18] | Ni1 | O2 | 90.334(8) |
| O1[4] | La1 | O3[7] | 60.7(2) | O2 | Ni1 | O2[17] | 180 |
| O1[1] | La1 | O3 | 60.7(2) | O2[18] | Ni1 | O2[17] | 89.666(8) |
| O1[2] | La1 | O3 | 60.5(2) | O2[8] | Ni1 | O2 | 89.666(7) |
| O1[3] | La1 | O3 | 119.7(2) | O2[8] | Ni1 | O2[18] | 180 |
| O1[2] | La1 | O3[6] | 119.7(2) | O2[8] | Ni1 | O2[17] | 90.334(8) |
| O1[4] | La1 | O3 | 119.5(2) | O1[2] | Ni2 | O4 | 180 |
| O1[2] | La1 | O3[7] | 60.5(2) | O3[17] | Ni2 | O1[2] | 90.1(2) |
| O1[4] | La1 | O3[6] | 119.5(2) | O3[19] | Ni2 | O1[2] | 90.1(2) |
| O1[4] | La1 | O3[5] | 60.7(2) | O3[7] | Ni2 | O1[2] | 90.1(2) |
| O1[1] | La1 | O3[7] | 119.5(2) | O3 | Ni2 | O1[2] | 90.1(2) |
| O1[3] | La1 | O3[7] | 119.7(2) | O3[19] | Ni2 | O3[17] | 89.666(8) |
| O2 | La1 | O1[1] | 59.8(2) | O3[19] | Ni2 | O3 | 90.334(8) |
| O2 | La1 | O1[2] | 59.6(2) | O3[7] | Ni2 | O3 | 89.666(8) |
| O2[8] | La1 | O1[3] | 120.2(2) | O3[19] | Ni2 | O3[7] | 179.7(5) |
| O2[9] | La1 | O1[2] | 120.2(2) | O3[7] | Ni2 | O3[17] | 90.334(8) |
| O2[5] | La1 | O1[1] | 120.0(2) | O3[17] | Ni2 | O3 | 179.7(5) |
| O2[9] | La1 | O1[1] | 59.8(2) | O3 | Ni2 | O4 | 89.9(2) |
| O2 | La1 | O1[3] | 120.2(2) | O3[17] | Ni2 | O4 | 89.9(2) |
| O2[5] | La1 | O1[4] | 59.8(2) | O3[19] | Ni2 | O4 | 89.9(2) |
| O2 | La1 | O1[4] | 120.0(2) | O3[7] | Ni2 | O4 | 89.9(2) |
| O2[5] | La1 | O1[3] | 59.6(2) | O5 | Ni3 | O5[17] | 180 |
| O2[5] | La1 | O1[2] | 120.2(2) | O5[24] | Ni3 | O5[17] | 90.334(8) |
| O2[8] | La1 | O1[4] | 59.8(2) | O5[24] | Ni3 | O5 | 89.666(8) |
| O2[9] | La1 | O1[3] | 59.6(2) | O5[14] | Ni3 | O5[17] | 89.666(7) |
| O2[8] | La1 | O1[2] | 59.6(2) | O5[24] | Ni3 | O5[14] | 180 |
| O2[9] | La1 | O1[4] | 120.0(2) | O5[14] | Ni3 | O5 | 90.334(8) |
| O2[8] | La1 | O1[1] | 120.0(2) | O5[14] | Ni3 | O6[20] | 90.000(2) |
| O2[5] | La1 | O2[8] | 60.555(18) | O5[14] | Ni3 | O6 | 90.000(2) |
| O2[8] | La1 | O2[9] | 90.63(3) | O5[17] | Ni3 | O6[20] | 90 |
| O2[5] | La1 | O2[9] | 60.166(18) | O5 | Ni3 | O6[20] | 90 |
| O2[8] | La1 | O2 | 60.166(18) | O5[24] | Ni3 | O6[20] | 90.000(2) |
| O2[5] | La1 | O2 | 90.63(3) | O5[17] | Ni3 | O6 | 90 |
| O2[9] | La1 | O2 | 60.555(18) | O5 | Ni3 | O6 | 90 |
| O2[5] | La1 | O3 | 178.85(12) | O5[24] | Ni3 | O6 | 90.000(2) |
| O2 | La1 | O3[7] | 120.11(7) | O6[20] | Ni3 | O6 | 180 |
| O2 | La1 | O3[6] | 120.49(7) | La1[4] | O1 | La1[3] | 90.000(2) |
| O2[9] | La1 | O3[6] | 90.53(11) | La1[4] | O1 | La1[2] | 90.000(1) |
| O2 | La1 | O3[5] | 178.85(12) | La1[1] | O1 | La1[2] | 90.000(2) |
| O2[5] | La1 | O3[6] | 120.11(7) | La1[1] | O1 | La1[3] | 90.000(1) |
| O2[9] | La1 | O3 | 120.49(7) | La1[1] | O1 | La1[4] | 179.8(6) |
| O2[8] | La1 | O3[5] | 120.49(7) | La1[3] | O1 | La1[2] | 179.8(6) |
| O2[8] | La1 | O3[7] | 90.53(11) | Ni1 | O1 | La1[3] | 90.1(3) |
| O2[8] | La1 | O3 | 120.11(7) | Ni1 | O1 | La1[2] | 90.1(3) |
| O2[5] | La1 | O3[7] | 120.49(7) | Ni1 | O1 | La1[1] | 90.1(3) |
| O2[5] | La1 | O3[5] | 90.53(11) | Ni1 | O1 | La1[4] | 90.1(3) |
| O2[8] | La1 | O3[6] | 178.85(12) | Ni1 | O1 | Ni2[2] | 180 |
| O2[9] | La1 | O3[5] | 120.11(7) | Ni2[2] | O1 | La1[2] | 89.9(3) |
| O2 | La1 | O3 | 90.53(11) | Ni2[2] | O1 | La1[3] | 89.9(3) |
| O2[9] | La1 | O3[7] | 178.85(12) | Ni2[2] | O1 | La1[4] | 89.9(3) |
| O3[6] | La1 | O3[7] | 88.3(2) | Ni2[2] | O1 | La1[1] | 89.9(3) |
| O3[7] | La1 | O3[5] | 59.22(14) | La1 | O2 | La1[12] | 90.63(3) |
| O3[6] | La1 | O3[5] | 58.84(13) | La1[3] | O2 | La1[1] | 90.62(3) |
| O3[5] | La1 | O3 | 88.3(2) | La1 | O2 | La1[3] | 89.37(3) |
| O3[7] | La1 | O3 | 58.84(13) | La1 | O2 | La1[1] | 180.00(3) |
| O3[6] | La1 | O3 | 59.22(13) | La1[12] | O2 | La1[3] | 180 |
| O3[5] | La2 | O3[7] | 64.71(15) | La1[12] | O2 | La1[1] | 89.38(3) |
| O3[6] | La2 | O3[7] | 98.0(3) | Ni1 | O2 | La1[1] | 89.763(5) |
| O3 | La2 | O3[6] | 64.72(15) | Ni1 | O2 | La1[12] | 89.763(6) |
| O3 | La2 | O3[7] | 64.29(15) | Ni1[10] | O2 | La1[3] | 89.763(6) |
| O3 | La2 | O3[5] | 98.0(3) | Ni1 | O2 | La1 | 90.237(5) |
| O3[5] | La2 | O3[6] | 64.29(15) | Ni1[10] | O2 | La1[1] | 90.237(5) |
| O3[6] | La2 | O4[11] | 65.4(2) | Ni1[10] | O2 | La1[12] | 90.237(6) |
| O3 | La2 | O4[11] | 129.67(19) | Ni1[10] | O2 | La1 | 89.763(5) |
| O3[5] | La2 | O4 | 129.67(19) | Ni1 | O2 | La1[3] | 90.237(5) |
| O3[7] | La2 | O4[5] | 65.6(2) | Ni1[10] | O2 | Ni1 | 180 |
| O3[7] | La2 | O4 | 65.4(2) | La1[12] | O3 | La1 | 88.3(2) |
| O3[5] | La2 | O4[5] | 65.6(2) | La2 | O3 | La1 | 86.84(3) |
| O3[5] | La2 | O4[10] | 129.48(19) | La2[12] | O3 | La1[12] | 86.84(3) |
| O3[7] | La2 | O4[11] | 129.67(19) | La2[12] | O3 | La1 | 175.2(2) |
| O3[7] | La2 | O4[10] | 129.48(19) | La2 | O3 | La1[12] | 175.2(2) |
| O3[5] | La2 | O4[11] | 65.4(2) | La2 | O3 | La2[12] | 98.0(3) |
| O3[6] | La2 | O4[10] | 65.6(2) | Ni2 | O3 | La1[12] | 89.67(17) |
| O3 | La2 | O4[5] | 129.48(19) | Ni2[10] | O3 | La1[12] | 90.13(18) |
| O3[6] | La2 | O4 | 129.67(19) | Ni2[10] | O3 | La1 | 89.67(17) |
| O3[6] | La2 | O4[5] | 129.48(19) | Ni2 | O3 | La1 | 90.13(18) |
| O3 | La2 | O4[10] | 65.6(2) | Ni2[10] | O3 | La2 | 89.84(16) |
| O3 | La2 | O4 | 65.4(2) | Ni2 | O3 | La2[12] | 89.84(16) |
| O4[5] | La2 | O4[11] | 88.35(9) | Ni2[10] | O3 | La2[12] | 90.34(16) |
| O4[11] | La2 | O4 | 160.5(5) | Ni2 | O3 | La2 | 90.34(16) |
| O4[5] | La2 | O4 | 88.35(8) | Ni2 | O3 | Ni2[10] | 179.7(5) |
| O4[10] | La2 | O4 | 88.35(8) | La2[16] | O4 | La2 | 160.5(5) |
| O4[10] | La2 | O4[11] | 88.35(8) | La2[17] | O4 | La2 | 88.35(8) |
| O4[10] | La2 | O4[5] | 160.4(5) | La2[17] | O4 | La2[12] | 160.4(5) |
| O6[10] | La2 | O3[7] | 131.00(13) | La2[17] | O4 | La2[16] | 88.35(8) |
| O6[10] | La2 | O3[6] | 131.00(13) | La2[12] | O4 | La2 | 88.35(8) |
| O6[10] | La2 | O3[5] | 131.00(14) | La2[12] | O4 | La2[16] | 88.35(8) |
| O6[10] | La2 | O3 | 131.00(14) | La3 | O4 | La2[12] | 99.8(2) |
| O6[10] | La2 | O4[11] | 80.3(2) | La3 | O4 | La2[16] | 99.7(2) |
| O6[10] | La2 | O4 | 80.3(2) | La3 | O4 | La2[17] | 99.8(2) |
| O6[10] | La2 | O4[10] | 80.2(3) | La3 | O4 | La2 | 99.7(2) |
| O6[10] | La2 | O4[5] | 80.2(3) | Ni2 | O4 | La2 | 80.3(2) |
| O4 | La3 | O5 | 132.463(16) | Ni2 | O4 | La2[17] | 80.2(2) |
| O4 | La3 | O5[12] | 132.463(16) | Ni2 | O4 | La2[12] | 80.2(2) |
| O4 | La3 | O5[14] | 132.463(16) | Ni2 | O4 | La2[16] | 80.3(2) |
| O4 | La3 | O5[15] | 132.463(16) | Ni2 | O4 | La3 | 180 |
| O4 | La3 | O6[10] | 80.7(2) | La3[21] | O5 | La3 | 84.93(3) |
| O4 | La3 | O6 | 80.7(2) | La3[5] | O5 | La3[25] | 84.93(3) |
| O4 | La3 | O6[13] | 80.7(2) | La3[21] | O5 | La3[25] | 95.07(3) |
| O4 | La3 | O6[12] | 80.7(2) | La3 | O5 | La3[25] | 180 |
| O5[12] | La3 | O5 | 95.07(3) | La3[5] | O5 | La3 | 95.07(3) |
| O5[12] | La3 | O5[14] | 62.681(18) | La3[21] | O5 | La3[5] | 180 |
| O5[15] | La3 | O5 | 62.681(18) | Ni3[10] | O5 | La3 | 90.246(6) |
| O5[12] | La3 | O5[15] | 63.089(18) | Ni3[10] | O5 | La3[5] | 89.754(6) |
| O5[15] | La3 | O5[14] | 95.07(3) | Ni3 | O5 | La3[25] | 90.246(6) |
| O5[14] | La3 | O5 | 63.089(18) | Ni3[10] | O5 | La3[25] | 89.754(6) |
| O5[12] | La3 | O6[10] | 128.69(18) | Ni3 | O5 | La3[21] | 89.754(6) |
| O5 | La3 | O6[13] | 128.51(18) | Ni3 | O5 | La3[5] | 90.246(6) |
| O5 | La3 | O6[10] | 65.9(2) | Ni3 | O5 | La3 | 89.754(6) |
| O5[15] | La3 | O6[10] | 65.95(19) | Ni3[10] | O5 | La3[21] | 90.246(6) |
| O5[15] | La3 | O6 | 128.51(18) | Ni3[10] | O5 | Ni3 | 180 |
| O5[14] | La3 | O6[10] | 128.69(18) | La2[17] | O6 | La3[5] | 99.3(2) |
| O5[15] | La3 | O6[12] | 128.69(18) | La2[17] | O6 | La3 | 99.3(2) |
| O5[12] | La3 | O6[12] | 65.9(2) | La2[17] | O6 | La3[17] | 99.3(2) |
| O5 | La3 | O6[12] | 128.69(18) | La2[17] | O6 | La3[23] | 99.3(2) |
| O5[15] | La3 | O6[13] | 66.2(2) | La3 | O6 | La3[23] | 161.4(5) |
| O5[12] | La3 | O6 | 128.51(18) | La3[17] | O6 | La3[5] | 161.5(5) |
| O5 | La3 | O6 | 66.2(2) | La3 | O6 | La3[17] | 88.50(8) |
| O5[12] | La3 | O6[13] | 66.2(2) | La3[23] | O6 | La3[5] | 88.50(8) |
| O5[14] | La3 | O6[12] | 65.95(19) | La3 | O6 | La3[5] | 88.50(8) |
| O5[14] | La3 | O6[13] | 128.51(18) | La3[23] | O6 | La3[17] | 88.50(8) |
| O5[14] | La3 | O6 | 66.2(2) | Ni3 | O6 | La2[17] | 180 |
| O6[13] | La3 | O6[10] | 88.51(8) | Ni3 | O6 | La3 | 80.7(2) |
| O6[13] | La3 | O6[12] | 88.51(8) | Ni3 | O6 | La3[17] | 80.7(2) |
| O6[13] | La3 | O6 | 161.4(5) | Ni3 | O6 | La3[23] | 80.7(2) |
| O6[10] | La3 | O6[12] | 161.5(5) | Ni3 | O6 | La3[5] | 80.7(2) |
| O6 | La3 | O6[10] | 88.50(8) | | | | |

[1] 3/2-X,1/2-Y,1-Z; [2] 1-X,1-Y,1-Z; [3] 1-X,-Y,1-Z; [4] 1/2-X,1/2-Y,1-Z; [5] -1/2+X,-1/2+Y,+Z; [6] 3/2-X,-1/2+Y,+Z; [7] 1-X,+Y,+Z; [8] -1/2+X,1/2-Y,1-Z; [9] +X,-Y,1-Z; [10] 1/2+X,-1/2+Y,+Z; [11] +X,-1+Y,+Z; [12] 1/2+X,1/2+Y,+Z; [13] 1+X,+Y,+Z; [14] +X,1-Y,2-Z; [15] 1/2+X,1/2-Y,2-Z; [16] +X,1+Y,+Z; [17] -1/2+X,1/2+Y,+Z; [18] +X,1-Y,1-Z; [19] 3/2-X,1/2+Y,+Z; [20] -X,1-Y,2-Z; [21] 1-X,1-Y,2-Z; [22] 1/2-X,3/2-Y,2-Z; [23] -1+X,+Y,+Z; [24] -1/2+X,1/2-Y,2-Z; [25] 1/2-X,1/2-Y,2-Z

Table S6 Fractional atomic coordinates (×10⁴) and equivalent isotropic displacement parameters (Å²×10³) for LNO-2222 (*Cmcm* setting with $a$ = 20.517(2), $b$ = 5.4536(6) and $c$ = 5.4053(6)). $U_{eq}$ is defined as 1/3 of the trace of the orthogonalized $U_{ij}$ tensor.

| Atom | x | y | z | $U_{eq}$ |
|---|---|---|---|---|
| La1 | 5000 | 2494.9(6) | 7500 | 6.79(14) |
| La2 | 6799.1(2) | 7578.2(4) | 2500 | 5.89(13) |
| Ni1 | 5960.7(4) | 7476.3(10) | 7500 | 4.12(17) |
| O1 | 5000 | 7100(11) | 7500 | 12.1(11) |
| O2 | 5895.1(19) | 10000 | 5000 | 8.7(8) |
| O3 | 6045(2) | 5000 | 5000 | 11.0(8) |
| O4 | 7956(2) | 7165(8) | 2500 | 14.5(9) |

**Table S7** Anisotropic Displacement Parameters (Å$^2$×10$^3$) for LNO-2222 (*Cmcm* setting). The Anisotropic displacement factor exponent takes the form: $-2\pi^2[h^2a^{*2}U_{11}+2hka^*b^*U_{12}+\ldots]$.

| Atom | Wyck. | $U_{11}$ | $U_{22}$ | $U_{33}$ | $U_{23}$ | $U_{13}$ | $U_{12}$ |
|---|---|---|---|---|---|---|---|
| La1 | 4c | 6.7(3) | 6.7(2) | 7.0(2) | 0 | 0 | 0 |
| La2 | 8g | 5.8(2) | 5.35(17) | 6.48(18) | 0 | 0 | -0.42(9) |
| Ni1 | 8g | 5.9(3) | 3.1(3) | 3.4(3) | 0 | 0 | 0.26(18) |
| O1 | 4c | 2(2) | 17(2) | 18(3) | 0 | 0 | 0 |
| O2 | 8e | 17(2) | 3.5(16) | 5.8(19) | 2.0(13) | 0 | 0 |
| O3 | 8e | 24(2) | 2.8(15) | 6.5(18) | -1.5(13) | 0 | 0 |
| O4 | 8g | 7(2) | 16.9(19) | 20(2) | 0 | 0 | 1.2(14) |

**Table S8** Selected bond lengths for LNO-2222 (*Cmcm* setting).

| Atom | Atom | Length/Å | Atom | Atom | Length/Å |
|---|---|---|---|---|---|
| La1 | O1 | 2.512(6) | La2 | O3$^9$ | 2.489(3) |
| La1 | O1$^1$ | 2.942(6) | La2 | O3 | 2.489(3) |
| La1 | O1$^2$ | 2.7116(6) | La2 | O4$^{10}$ | 2.7524(9) |
| La1 | O1$^3$ | 2.7116(6) | La2 | O4$^{11}$ | 2.7524(9) |
| La1 | O2$^1$ | 2.655(3) | La2 | O4$^{12}$ | 2.995(4) |
| La1 | O2$^4$ | 2.655(3) | La2 | O4$^{13}$ | 2.551(4) |
| La1 | O2$^5$ | 2.655(3) | La2 | O4 | 2.385(4) |
| La1 | O2$^2$ | 2.655(3) | Ni1 | O1 | 1.9816(11) |
| La1 | O3$^5$ | 2.880(3) | Ni1 | O2$^6$ | 1.9335(5) |
| La1 | O3$^6$ | 2.880(3) | Ni1 | O2 | 1.9335(5) |
| La1 | O3 | 2.880(3) | Ni1 | O3$^6$ | 1.9183(6) |
| La1 | O3$^2$ | 2.880(3) | Ni1 | O3 | 1.9183(6) |
| La2 | O2 | 2.648(3) | Ni1 | O4$^{10}$ | 2.231(4) |
| La2 | O2$^9$ | 2.648(3) | | | |

$^1$+X,-1+Y,+Z; $^2$1-X,1-Y,1-Z; $^3$1-X,1-Y,2-Z; $^4$+X,-1+Y,3/2-Z; $^5$1-X,1-Y,1/2+Z; $^6$+X,+Y,3/2-Z; $^7$+X,2-Y,-1/2+Z; $^8$+X,+Y,-1+Z; $^9$+X,+Y,1/2-Z; $^{10}$3/2-X,3/2-Y,1-Z; $^{11}$3/2-X,3/2-Y,-Z; $^{12}$3/2-X,-1/2+Y,1/2-Z; $^{13}$3/2-X,1/2+Y,1/2-Z; $^{14}$+X,+Y,1+Z; $^{15}$+X,1-Y,1/2+Z; $^{16}$+X,2-Y,1/2+Z; $^{17}$+X,1+Y,+Z; $^{18}$1-X,+Y,3/2-Z; $^{19}$+X,1-Y,-1/2+Z

**Table S9** Selected bond angles for LNO-2222 (*Cmcm* setting).

| Atom | Atom | Atom | Angle/° | Atom | Atom | Atom | Angle/° |
|---|---|---|---|---|---|---|---|
| O1 | La1 | O1$^1$ | 85.33(12) | O4 | La2 | O2 | 138.11(9) |
| O1 | La1 | O1$^2$ | 180 | O4 | La2 | O3 | 124.41(11) |
| O1 | La1 | O1$^3$ | 85.33(12) | O4 | La2 | O3$^9$ | 124.41(11) |
| O1$^1$ | La1 | O1$^2$ | 94.67(12) | O4$^{10}$ | La2 | O4$^{11}$ | 158.17(18) |
| O1$^3$ | La1 | O1$^2$ | 94.67(12) | O4 | La2 | O4$^{11}$ | 79.82(9) |
| O1$^1$ | La1 | O1$^3$ | 170.7(2) | O4 | La2 | O4$^{13}$ | 84.08(12) |
| O1 | La1 | O2$^4$ | 120.82(3) | O4$^{13}$ | La2 | O4$^{10}$ | 85.08(10) |
| O1 | La1 | O2$^2$ | 120.82(3) | O4 | La2 | O4$^{10}$ | 79.82(9) |
| O1 | La1 | O2$^5$ | 120.82(3) | O4$^{13}$ | La2 | O4$^{12}$ | 159.00(18) |
| O1 | La1 | O2$^3$ | 120.82(3) | O4 | La2 | O4$^{12}$ | 74.93(13) |
| O1$^1$ | La1 | O3$^6$ | 59.58(7) | O4$^{13}$ | La2 | O4$^{11}$ | 85.08(10) |
| O1$^3$ | La1 | O3$^6$ | 115.41(8) | O4$^{11}$ | La2 | O4$^{12}$ | 91.12(9) |
| O1$^1$ | La1 | O3$^3$ | 115.41(8) | O4$^{10}$ | La2 | O4$^{12}$ | 91.12(9) |
| O1$^1$ | La1 | O3$^4$ | 59.58(7) | O1 | Ni1 | O4$^{10}$ | 179.09(19) |

| | | | | | | | | |
|---|---|---|---|---|---|---|---|---|
| O1 | La1 | O3[6] | 61.68(4) | | O2[6] | Ni1 | O1 | 90.26(17) |
| O1 | La1 | O3[4] | 61.68(4) | | O2 | Ni1 | O1 | 90.26(17) |
| O1[3] | La1 | O3[3] | 59.58(7) | | O2[6] | Ni1 | O2 | 88.68(3) |
| O1 | La1 | O3 | 61.68(4) | | O2[6] | Ni1 | O4[10] | 90.39(14) |
| O1 | La1 | O3[3] | 61.68(4) | | O2 | Ni1 | O4[10] | 90.39(14) |
| O1[3] | La1 | O3 | 59.58(7) | | O3[6] | Ni1 | O1 | 90.99(18) |
| O1[3] | La1 | O3[4] | 115.41(8) | | O3 | Ni1 | O1 | 90.99(18) |
| O1[1] | La1 | O3 | 115.41(8) | | O3 | Ni1 | O2[6] | 178.67(17) |
| O2[5] | La1 | O1[1] | 62.26(8) | | O3 | Ni1 | O2 | 90.865(16) |
| O2[4] | La1 | O1[3] | 123.29(8) | | O3[6] | Ni1 | O2 | 178.68(17) |
| O2[4] | La1 | O1[2] | 59.18(3) | | O3[6] | Ni1 | O2[6] | 90.864(16) |
| O2[2] | La1 | O1[2] | 59.18(3) | | O3[6] | Ni1 | O3 | 89.57(3) |
| O2[5] | La1 | O1[2] | 59.18(3) | | O3[6] | Ni1 | O4[10] | 88.37(15) |
| O2[3] | La1 | O1[2] | 59.18(3) | | O3 | Ni1 | O4[10] | 88.37(15) |
| O2[3] | La1 | O1[1] | 123.29(8) | | La1[3] | O1 | La1[17] | 85.33(12) |
| O2[4] | La1 | O1[1] | 62.26(8) | | La1[1] | O1 | La1[17] | 85.33(12) |
| O2[3] | La1 | O1[3] | 62.26(8) | | La1 | O1 | La1[1] | 94.67(12) |
| O2[2] | La1 | O1[3] | 62.26(8) | | La1 | O1 | La1[17] | 180 |
| O2[2] | La1 | O1[1] | 123.29(8) | | La1[1] | O1 | La1[3] | 170.7(2) |
| O2[5] | La1 | O1[3] | 123.29(8) | | La1 | O1 | La1[3] | 94.67(12) |
| O2[5] | La1 | O2[2] | 61.18(7) | | Ni1[18] | O1 | La1[1] | 89.52(3) |
| O2[3] | La1 | O2[2] | 87.53(12) | | Ni1 | O1 | La1[3] | 89.52(3) |
| O2[4] | La1 | O2[2] | 118.35(7) | | Ni1 | O1 | La1 | 95.94(17) |
| O2[3] | La1 | O2[5] | 118.35(7) | | Ni1[18] | O1 | La1[3] | 89.52(3) |
| O2[5] | La1 | O2[4] | 87.53(12) | | Ni1[18] | O1 | La1[17] | 84.06(17) |
| O2[3] | La1 | O2[4] | 61.18(7) | | Ni1 | O1 | La1[17] | 84.06(17) |
| O2[2] | La1 | O3[6] | 88.10(9) | | Ni1[18] | O1 | La1 | 95.94(17) |
| O2[2] | La1 | O3 | 59.28(5) | | Ni1 | O1 | La1[1] | 89.52(3) |
| O2[3] | La1 | O3 | 121.30(5) | | Ni1[18] | O1 | Ni1 | 168.1(3) |
| O2[4] | La1 | O3[4] | 59.28(5) | | La1[3] | O2 | La1[17] | 92.47(12) |
| O2[2] | La1 | O3[4] | 175.62(8) | | La2[14] | O2 | La1[17] | 88.231(11) |
| O2[4] | La1 | O3[6] | 121.29(5) | | La2 | O2 | La1[3] | 88.232(11) |
| O2[3] | La1 | O3[6] | 175.62(8) | | La2 | O2 | La1[17] | 179.07(9) |
| O2[3] | La1 | O3[4] | 88.10(9) | | La2[14] | O2 | La1[3] | 179.07(9) |
| O2[3] | La1 | O3[3] | 59.28(5) | | La2[14] | O2 | La2 | 91.07(12) |
| O2[5] | La1 | O3[3] | 175.62(8) | | Ni1[8] | O2 | La1[3] | 93.28(8) |
| O2[4] | La1 | O3[3] | 88.10(9) | | Ni1[8] | O2 | La1[17] | 92.24(8) |
| O2[2] | La1 | O3[3] | 121.29(5) | | Ni1 | O2 | La1[3] | 92.24(8) |
| O2[5] | La1 | O3 | 88.09(9) | | Ni1 | O2 | La1[17] | 93.28(8) |
| O2[5] | La1 | O3[4] | 121.29(5) | | Ni1 | O2 | La2[14] | 87.11(8) |
| O2[4] | La1 | O3 | 175.62(8) | | Ni1[8] | O2 | La2[14] | 87.30(9) |
| O2[5] | La1 | O3[6] | 59.28(5) | | Ni1[8] | O2 | La2 | 87.12(8) |
| O3[4] | La1 | O1[2] | 118.32(4) | | Ni1 | O2 | La2 | 87.30(9) |
| O3 | La1 | O1[2] | 118.32(4) | | Ni1 | O2 | Ni1[8] | 172.0(2) |
| O3[3] | La1 | O1[2] | 118.32(4) | | La1 | O3 | La1[3] | 83.72(12) |
| O3[6] | La1 | O1[2] | 118.32(4) | | La2 | O3 | La1[3] | 86.56(2) |
| O3[6] | La1 | O3[4] | 96.28(12) | | La2[16] | O3 | La1 | 86.56(2) |
| O3[6] | La1 | O3 | 55.97(7) | | La2[16] | O3 | La1[3] | 170.25(14) |
| O3[4] | La1 | O3 | 123.36(7) | | La2 | O3 | La1 | 170.25(14) |
| O3[6] | La1 | O3[3] | 123.36(7) | | La2[16] | O3 | La2 | 103.17(16) |
| O3[3] | La1 | O3 | 96.28(12) | | Ni1 | O3 | La1 | 86.33(10) |

| | | | | | | | |
|---|---|---|---|---|---|---|---|
| O3[3] | La1 | O3[4] | 55.97(7) | Ni1[19] | O3 | La1 | 85.94(10) |
| O2[9] | La2 | O2 | 61.38(7) | Ni1 | O3 | La1[3] | 85.94(10) |
| O2 | La2 | O4[10] | 66.50(10) | Ni1[19] | O3 | La1[3] | 86.33(10) |
| O2[9] | La2 | O4[10] | 127.12(9) | Ni1 | O3 | La2 | 92.35(8) |
| O2 | La2 | O4[11] | 127.12(9) | Ni1[19] | O3 | La2[16] | 92.35(8) |
| O2[9] | La2 | O4[11] | 66.50(10) | Ni1[19] | O3 | La2 | 94.10(8) |
| O2[9] | La2 | O4[12] | 127.53(7) | Ni1 | O3 | La2[16] | 94.10(8) |
| O2 | La2 | O4[12] | 127.53(7) | Ni1[19] | O3 | Ni1 | 169.6(3) |
| O3 | La2 | O2 | 64.50(6) | La2[11] | O4 | La2[13] | 85.37(9) |
| O3 | La2 | O2[9] | 97.11(10) | La2 | O4 | La2[10] | 100.18(9) |
| O3[9] | La2 | O2 | 97.10(10) | La2[12] | O4 | La2[10] | 90.80(9) |
| O3[9] | La2 | O2[9] | 64.50(6) | La2[10] | O4 | La2[11] | 158.17(18) |
| O3 | La2 | O3[9] | 65.77(8) | La2 | O4 | La2[13] | 94.23(13) |
| O3 | La2 | O4[12] | 63.08(8) | La2 | O4 | La2[12] | 106.77(16) |
| O3[9] | La2 | O4[13] | 132.54(7) | La2 | O4 | La2[11] | 100.18(9) |
| O3[9] | La2 | O4[10] | 132.47(10) | La2[12] | O4 | La2[11] | 90.80(9) |
| O3[9] | La2 | O4[12] | 63.08(8) | La2[12] | O4 | La2[13] | 159.00(18) |
| O3 | La2 | O4[11] | 132.47(10) | La2[10] | O4 | La2[13] | 85.37(9) |
| O3 | La2 | O4[10] | 66.98(10) | Ni1[10] | O4 | La2[13] | 75.32(12) |
| O3[9] | La2 | O4[11] | 66.98(10) | Ni1[10] | O4 | La2[10] | 79.27(9) |
| O3 | La2 | O4[13] | 132.54(7) | Ni1[10] | O4 | La2[11] | 79.27(9) |
| O4 | La2 | O2[9] | 138.11(9) | Ni1[10] | O4 | La2 | 169.5(2) |
| O4[13] | La2 | O2 | 69.44(9) | Ni1[10] | O4 | La2[12] | 83.68(14) |
| O4[13] | La2 | O2[9] | 69.44(9) | | | | |

[1]1-X,1-Y,2-Z; [2]+X,-1+Y,+Z; [3]1-X,1-Y,1-Z; [4]1-X,1-Y,1/2+Z; [5]+X,-1+Y,3/2-Z; [6]+X,+Y,3/2-Z; [7]+X,+Y,-1+Z; [8]+X,2-Y,-1/2+Z; [9]+X,+Y,1/2-Z; [10]3/2-X,3/2-Y,1-Z; [11]3/2-X,3/2-Y,-Z; [12]3/2-X,-1/2+Y,1/2-Z; [13]3/2-X,1/2+Y,1/2-Z; [14]+X,2-Y,1/2+Z; [15]+X,+Y,1+Z; [16]+X,1-Y,1/2+Z; [17]+X,1+Y,+Z; [18]1-X,+Y,3/2-Z; [19]+X,1-Y,-1/2+Z

**Table S10** Crystal data and structure refinement results for LNO-1313 with a putative *Imma* structure. Note that an alternative setting is *Imcm* with $a$ = 5.4361(6), $b$ = 5.4717(6) and $c$ = 40.713(4).

| Identification code | LNO-1313_Imma |
| --- | --- |
| Empirical formula | $La_3Ni_2O_7$ |
| Formula weight | 646.15 |
| Temperature/K | 295 |
| Crystal system | orthorhombic |
| Space group | *Imma* |
| $a$/Å | 40.713(4) |
| $b$/Å | 5.4361(6) |
| $c$/Å | 5.4717(6) |
| $\alpha$/° | 90 |
| $\beta$/° | 90 |
| $\gamma$/° | 90 |
| Volume/Å$^3$ | 1211.0(2) |
| Z | 8 |
| $\rho_{calc}$g/cm$^3$ | 7.088 |
| $\mu$/mm$^{-1}$ | 26.799 |
| F(000) | 2264.0 |
| Crystal size/mm$^3$ | 0.101 × 0.081 × 0.014 |
| Radiation | MoK$\alpha$ ($\lambda$ = 0.71073) |
| 2$\Theta$ range for data collection/° | 2 to 60.988 |
| Index ranges | -58 ≤ h ≤ 58, -7 ≤ k ≤ 7, -7 ≤ l ≤ 7 |
| Reflections collected | 13421 |
| Independent reflections | 1029 [$R_{int}$ = 0.0614, $R_{sigma}$ = 0.0341] |
| Refinement method | Full-matrix least-squares on $F^2$ |
| Absorption correction | Numerical |
| Goodness-of-fit on $F^2$ | 1.236 |
| Final $R$ indexes [$I$>=2$\sigma$ ($I$)] | $R_1$ = 0.0623, $wR_2$ = 0.1496 |
| Final $R$ indexes [all data] | $R_1$ = 0.0702, $wR_2$ = 0.1522 |
| Largest diff. peak/hole / e Å$^{-3}$ | 3.23/-4.36 |

**Table S11** Fractional atomic coordinates (×10$^4$) and equivalent isotropic displacement parameters (Å$^2$×10$^3$) for LNO-1313 with a putative *Imma* structure. $U_{eq}$ is defined as 1/3 of the trace of the orthogonalized $U_{ij}$ tensor.

| Atom | Wyck. | x | y | z | U(eq) |
|---|---|---|---|---|---|
| La1 | 8*i* | 2032.7(3) | -2500 | 2495(3) | 9.2(3) |
| La2 | 8*i* | 1132.1(3) | -2500 | 2551(2) | 5.8(3) |
| La3 | 8*i* | 434.0(3) | 2500 | 2480(2) | 7.1(3) |
| Ni1 | 4*c* | 2500 | -7500 | 2500 | 3.4(7) |
| Ni2 | 8*i* | 1545.5(7) | -7500 | 2486(5) | 4.8(5) |
| Ni3 | 4*e* | 0 | -2500 | 2512(8) | 5.5(7) |
| O1 | 8*i* | 2027(4) | -7500 | 2040(30) | 9(3) |
| O2 | 8*f* | 2544(4) | -10000 | 0 | 26(5) |
| O3 | 8*f* | 1576(4) | -5000 | 5000 | 7(3) |
| O4 | 8*f* | 1513(4) | -10000 | 0 | 5(3) |
| O5 | 8*i* | 1008(4) | -7500 | 2720(40) | 12(3) |
| O6 | 8*i* | 549(6) | -2500 | 2390(50) | 35(6) |
| O7 | 4*b* | 0 | -5000 | 5000 | 31(9) |
| O8 | 4*a* | 0 | 0 | 0 | 21(7) |